\documentclass[preprintnumbers, onecolumn, floatfix, preprintnumbers, amsmath, amssymb, superscriptaddress]{revtex4}

\usepackage[colorlinks,linkcolor=red,urlcolor=blue,citecolor=blue]{hyperref}
\usepackage{amssymb, amsmath, bm, dcolumn, epsf, graphicx, latexsym, mathbbol, slashed}
\usepackage{mathrsfs}
\usepackage{comment}
\usepackage{color}

\newcommand{\dd}[0]{\mathrm{d}}

\begin{document}

\title{Gravitational Radiation from Eccentric Binary Black Hole System\\ in Dynamical Chern-Simons Gravity}

\author{Zhao Li}
\email{lz111301@mail.ustc.edu.cn}
\affiliation{ Department of Astronomy, University of Science and Technology of China, Hefei, Anhui 230026, China,\\ and School of Astronomy and Space Science, University of Science and Technology of China, Hefei 230026, China}
\author{Jin Qiao}
\affiliation{School of Fundamental Physics and Mathematical Sciences, Hangzhou Institute for Advanced Study, UCAS, Hangzhou 310024, China}
\affiliation{University of Chinese Academy of Sciences, 100049/100190 Beijing, China}
\author{Tan Liu}
\affiliation{School of Fundamental Physics and Mathematical Sciences, Hangzhou Institute for Advanced Study, UCAS, Hangzhou 310024, China}
\affiliation{University of Chinese Academy of Sciences, 100049/100190 Beijing, China}
\author{Rui Niu}
\affiliation{ Department of Astronomy, University of Science and Technology of China, Hefei, Anhui 230026, China,\\ and School of Astronomy and Space Science, University of Science and Technology of China, Hefei 230026, China}
\author{Shaoqi Hou}
\affiliation{School of Physics and Technology, Wuhan University, Wuhan, Hubei 430072, China}
\author{Tao Zhu}
\affiliation{Institute for Theoretical Physics and Cosmology, Zhejiang University of Technology, Hangzhou, 310032, China,\\
 United Center for Gravitational Wave Physics (UCGWP), Zhejiang University of Technology, Hangzhou, 310032, China}
\author{Wen Zhao}
\email{wzhao7@ustc.edu.cn}
\affiliation{ Department of Astronomy, University of Science and Technology of China, Hefei, Anhui 230026, China,\\ and School of Astronomy and Space Science, University of Science and Technology of China, Hefei 230026, China}

\begin{abstract}
Dynamical Chern-Simons (DCS) gravity, a typical parity-violating gravitational theory, modifies both the generation and propagation of gravitational waves from general relativity (GR). In this work, we derive the gravitational waveform radiated from a binary slowly-rotating black hole system with eccentric orbits under the spin-aligned assumption in the DCS theory. Compared with GR, DCS modification enters the second-order post-Newtonian (2PN) approximation, affecting the spin-spin coupling and monopole-quadrupole coupling of binary motion. This modification produces an extra precession rate of periastron. This effect modulates the scalar and gravitational waveform through a quite low frequency. Additionally, the dissipation of conserved quantities results in the secular evolution of the semimajor axis and the eccentricity of binary orbits. Finally, the frequency-domain waveform is given in the post-circular scheme, requiring the initial eccentricity to be $\lesssim0.3$. This ready-to-use template will benefit the signal searches and improve the future constraint on DCS theory.
\end{abstract}

\maketitle

\section{\label{Introduction}Introduction}

General relativity (GR) is always considered the most successful theory of gravity \cite{Chandrasekhar1984}. However, various difficulties of this theory are also well known. On the theoretical side, GR has singularity and quantization problems \cite{Penrose1965, Dewitt1967, Kiefer2007}. On the experimental side, all the observations in the cosmological scale indicate the existence of so-called dark matter and dark energy \cite{Sahni2005, Spergel2015, Peebles2003, Oks2021, Bertone2018}, which might mean that GR is invalid at this scale. For these reasons, we now must experimentally test GR in a variety of different spacetime environments and astrophysical scales. Since then, the gravitational tests on the submillimeter scale \cite{Hoyle2001, Sabulsky2019}, in the solar system \cite{Weinberg, Will2014, Will2014test, GPB2011, Shapiro1990, Smith2008, Marchi2020}, in the binary-pulsar systems \cite{HulseTaylor1975, Stairs2003, Kramer2006, Hu2023, Yunes2009CStest}, and in the astrophysical and cosmological scales \cite{Berti2015, Ishak2019, Gralla2021, Uzan2011} have been found to agree remarkably with Einstein’s theory.

The direct observation of gravitational waves (GWs) provided a new probe to test gravity in extreme-gravity environments. As predicted by GR, the currently observable GW can only be generated in strong gravitational fields and hardly interacts with matter, carrying information about the nature of gravity in the strong-field regime. In recent years, Laser Interferometer Gravitational-wave Observatory (LIGO) and Virgo collaboration have detected 90 GW signals radiated from compact binary coalescence events \cite{GWTC1, GWTC2, GWTC21, GWTC3}, for example, the well-known GW150914 \cite{GW150914} and GW190521 \cite{GW190521} from binary black holes (BBH) system, GW170817 \cite{GW170817} from binary neutron stars system, and GW200105 and GW200115 from neutron stars-black hole system \cite{GW200105andGW200115}. More information about these identified signals can be found in Refs.\,\cite{GWTC1, GWTC2, GWTC21, GWTC3}. Numerous works have used GW data to perform gravitational tests \cite{GW150914test, GW170817test, GWTC1test, GWTC2test, GWTC3test, Carson2020, Krishnendu2021, Ghosh2022, Puecher2022, Chamberlain2017, Barausse2016, Silva2021}. Currently, the third-generation GW observatories,  Einstein Telescope (ET) \cite{Punturo2010} and Cosmic Explorer (CE) \cite{Reitze2019}, as well as the space-based detectors, e.g., LISA \cite{Robson2019}, TianQin \cite{Luo2016} and Taiji \cite{ZirenLuo2020,ZirenLuo2021}, are under construction. These detectors are expected to observe several different astrophysical GW events, among which the compact binary inspirals are still one of the most promising sources \cite{Broeck2014,Salvatore2017,Chan2018}.

This work focuses on the BBH systems and the radiated GWs during the process of inspiraling to merging \cite{Will2014,MTW,Maggiore2008}. Because of the extremely strong gravitational field near the binary system in the pre-merger phase, the radiated GW may encode the distinction between the alternative gravitational theory and GR. This is why the detection of GWs generated from BBH is an essential aspect of the gravitational experiment \cite{YunesppE2009,Li2012,Agathos2014}. However, for the detection of GW signals and subsequent extraction of physical parameters from them, a set of precise theoretical templates is required \cite{Jaranowski2012}. GW events will be confirmed only when a sufficiently high signal-to-noise ratio (SNR) of the signal with the template is achieved. Thus, the templates have to be highly accurate, because a systematic inaccuracy can underestimate the SNR and lead to a missed detection. However, these templates for highly nonlinear and strongly general relativistic processes, e.g., the evolution of BBH systems, can be constructed only by numerical-relativity simulations \cite{Palenzuela2020,Shibata2015}, which is computationally expensive, especially for modeling the theories beyond GR \cite{Okounkova2020a,Okounkova2020b,Okounkova2023}. The post-Newtonian (PN) approximation \cite{EIH,Blanchet2014} is an alternative method to model the gravitational waveform from BBH in the pre-merger phase, where the separation between two bodies is much larger than the gravitational radii of them and, equivalently, the relative velocity is much smaller than the speed of light in the vacuum, i.e., $v^2\sim m/r\ll1$. In this framework, thus, the bodies can be regarded effectively as point particles \cite{EIH,Blanchet2014}. Until recently, the analytical PN expansion was known for non-spinning systems up to the 3.5PN, i.e., up to the $v^7$ correction beyond the leading-order quadrupole formula \cite{Maggiore2008,Blanchet2014,Blanchet2002b,Arun2005,Faye2006}. Refs.\,\cite{Marchand2020,Larrouturou2022,Blanchet2022,Blanchet2023a,Blanchet2023b} have pushed the accuracy to the next 4PN level. Adopting such an approach, Refs.\,\cite{Maggiore2008,Will2014} gave the templates in Newtonian order for non-spinning BBH systems with circular orbits, or quasi-circular orbits, more accurately, due to the radiation reaction. And the higher-order corrections can be found in Refs.\,\cite{Blanchet1995,Arun2004,Blanchet2002a,Blanchet2008}. 

Quasi-circular orbits are a reasonable choice because the GW dissipation circularizes the orbits of isolated BBH with an initial eccentricity, by the time their orbital frequency reaches the sensitive bands of ground-based GW observatories \cite{PetersMathews1963}. However, recent analyses \cite{Romero-Shaw2020,Gayathri2022,Romero-Shaw2022} found evidence of eccentricity in GW150921 and several other events, indicating these binary systems formed dynamically in densely populated environments. Ignoring eccentricity will result in an illusion of a deviation from GR \cite{Pankaj2022,Narayan2023,Sajad2023}. The templates for eccentric binary are also an essential topic in both numerical simulations \cite{ZhoujianCao2017} and analytical calculations \cite{PetersMathews1963,Yunes2009ecc}. When regarding the two bodies as point masses in the PN framework, the bound orbits are accurately elliptic, parameterized by semimajor axis and eccentricity, at the Newtonian-order approximation. In this simplest case, Refs.\,\cite{PetersMathews1963,Damour1983} gave the energy dissipation, and Yunes et al. \cite{Yunes2009ecc} obtained the analytic expression for such a frequency-domain template in the small-eccentricity limit. However, the higher-order effects bring some difficulties in getting the templates. On the one hand, starting from the 1PN approximation, such correction induces the well-known periastron advance effect \cite{Will2014}, causing the binary orbit to be no longer closed. To solve this problem, the quasi-Kerperian (QK) parameterization is usually adopted to integrate the equations of motion (EOM). The results for non-spinning BBH systems were derived by Refs.\,\cite{Damour1985,Will2014} at 1PN order, by Refs.\,\cite{Damour1988,Wex1995} at 2PN order, and by Ref.\,\cite{Memmesheimer2004} at 3PN order. Based on the precise descriptions of binary motions, the gravitational waveforms are logically obtained \cite{Moreno-Garrido1995,Gopakumar1997,Gopakumar2002,Damour2004,Arun2008,Konigsdorffer2006}, where the periastron advance modulates the waveforms with a much lower frequency. On the other hand, starting from the 1.5PN approximation, the non-aligned spins of the objects influence the motion via spin-orbit (SO), spin-spin (SS), and monopole-quadrupole (MQ) couplings \cite{Blanchet2006,Bohe2013,Bohe2015,Buonanno2013,Kidder1995,Cho2021}. In particular, the spins' components, perpendicular to the orbit, produce the orbital precession \cite{Chamberlain2017} and prevent the orbital circularization during energy dissipation, disappearing for the spin-aligned case \cite{Klein2010,Klein2018}. The efforts involving the spin effects can be found in Refs.\,\cite{Tessmer2010a,Klein2010,Gopakumar2011,Csizmadia2012,Tessmer2013}. The corresponding frequency-domain waveform can be found in Refs.\,\cite{Tessmer2010b,Loutrel2019,Loutrel2020,Klein2018,Tanay2016,Moore2016,Moore2019,Tiwari2019}.

The GW-based gravitational test entails comparing the signals predicted by GR and those by the alternative theories and constraining their differences by observations \cite{Yunes2010PVtest,XingZhangNSMGtest,RuiNiu2019,RuiNiu2020,RuiNiu2021test,ChangfuShi2022,Carson2020b,Chamberlain2017,Barausse2016,Alexander2018,Perkins2021,Perkins2021,WenZhao2020test,Mirshekari2012,QiangWu2022,JinQiao2023,ChengGong2022,YifanWang2022,TaoZhu2022,ChangfuShi2022,Nair2019,Perkins2021}, in which the modified templates via PN approximation are required. Modified gravitational waveforms have been derived in various extended gravitational theories for quasi-circular \cite{XingZhang2017BD,TanLiu2020BD,XiangZhao2019,ChaoZhang2020,KaiLin2019,XingZhang2016ppNSMG,XingZhang2017SMG,TanLiu2018SMG} and quasi-elliptic \cite{XingZhang2019SMG2,Chowdhuri2022,Vittorio2023} binary systems at the leading Newtonian order, that is the leading-order beyond-GR modification and the sub-leading effects are negligible. Unlike the above theories, as a kind of parity-violating theory \cite{MingzheLi2020NY,MingzheLi2021NY,MingzheLi2022STG,JinQiao2023}, dynamical Chern-Simons (DCS) gravity \cite{ChernSimons2003,Alexander2009} modifies the waveform from BBH at 2PN approximation, leaving behind a lower-order waveform that is completely consistent with GR \cite{Loutrel2018,Yagi2012pn,Yagi2012gw,ZhaoLi2023}. The template under quasi-circular and spin-aligned assumption has been reported in our previous works \cite{Yagi2012gw,ZhaoLi2023}. Then, the quasi-Keplerian case is taken into consideration in this paper, including the periastron advance but still assuming the spin vectors to be aligned with the orbital angular momentum (OAM). So the spin precession effects \cite{Loutrel2022} exceed the range of this article.

This work aims to derive the quasi-Keplerian motion, the time-domain gravitational waveform, orbital evolution due to the radiation reaction, and the frequency-domain waveform for the non-precessing BBH in DCS gravity. Because of the spin-aligned assumption, the bodies are always in the orbital plane. So the quasi-Keplerian parameterization \cite{Damour1983,Damour1985,Will2014,Wex1995,Memmesheimer2004} is successfully applied to the DCS extension, involving two elements, the ``radial" semimajor axis $a_r$ (equivalently the orbital frequency $F$ or its dimensionless version $x$) and ``radial" eccentricity $e_r$. The final result presents a doubly periodic structure and predicts the precession rate of the periastron. Precession makes the azimuth angle to be no longer a suitable periodic variable. Two alternative angular variables, the true anomaly, and the mean anomaly are used to express the waveform. Both of these versions show the low-frequency modulation from the precession effect. Based on the motion and time-domain waveform, the dissipation rate of energy and orbital angular momentum are obtained and we can get the secular evolution of elements. The orbital circularization is described by equation $dx/de_r$, with analytical solution $x=x(e_r)$. Although the extra scalar field slightly affects the rate of circularization, the eccentric orbits are still finally reduced to be quasi-circular through radiation reaction. To finally calculate the waveform in Fourier space, the eccentricity is written as a function of frequency, $e_r(F)$, up to $\mathcal{O}(e_r^4)$ order, valid for $e_0\lesssim0.3$. Therefore, the stationary phase approximation (SPA) \cite{Tessmer2010b,Klein2018} gives the Fourier transformation of the time-domain waveform and the ready-to-use template. These results will benefit the signal searches and improve the theoretical constraints of DCS theory in the future.

This paper is organized as follows. In Section \ref{sec:DCS}, we briefly review the DCS theory. In Section \ref{sec:motion}, we give conserved quantities, EOM, and its QK parameterization solution up to the leading-order DCS modification. Section \ref{sec:waveform} calculates the scalar and tensor gravitational waveforms and their polarizations. Furthermore, the energy, angular-momentum fluxes carried by radiation, and the secular evolutions of QK orbital elements are presented in Section \ref{sec:radiation-reaction}. The post-circular frequency-domain waveform is shown in \ref{sec:Fourier-waveform}. Finally, we make a summary and discussion in Section \ref{sec:conslusion}. Some complicated but unimportant modified coefficients are listed in Appendices \ref{Appendix-A}, \ref{Appendix-B}, \ref{Appendix-C}, \ref{Appendix-D}, and \ref{Appendix-E}. Throughout the paper, we work in geometric units in which $c=G=1$, where $c$ is the speed of light in the vacuum and $G$ is the gravitational constant.

\section{\label{sec:DCS}DCS Gravity}
In this section, we outline the basic knowledge of DCS theory \cite{ChernSimons2003,Alexander2009}. The full action of the DCS theory is
\begin{equation}
\label{action}
S=\int\dd^4x\sqrt{-g}\left[\frac{1}{16\pi}R+\frac{\alpha}{4}\vartheta R_{\nu\mu\alpha\beta}\hat{R}^{\mu\nu\alpha\beta}
-\frac{\beta_0}{2}(\nabla_{\mu}\vartheta)(\nabla^{\mu}\vartheta)+\mathcal{L}_{m}\right],
\end{equation}
where the gravity is described by a pseudo scalar field $\vartheta$ and the metric $g_{\mu\nu}$. The first term in Eq.\,(\ref{action}) gives the Einstein-Hilbert action, where $g$ is the determinant of the metric $g_{\mu\nu}$ and $R$ is the Ricci scalar. The second term describes the coupling between a pseudo scalar field ($\vartheta$) and the Pontryagin density ($R_{\nu\mu\alpha\beta}\hat{R}^{\mu\nu\alpha\beta}$) which is a kind of parity-violating modification, causing the non-conservation of the DCS topological current. $\alpha$ is the coupling parameter. $R_{\nu\mu\rho\sigma}$ is Riemann tensor and its dual is defined as $\hat{R}^{\mu\nu\alpha\beta}\equiv(1/2)\varepsilon^{\rho\sigma\alpha\beta}R^{\mu\nu}_{\ \ \alpha\beta}$, with $\varepsilon^{\rho\sigma\alpha\beta}$ being the Levi-Civit\'{a} tensor defined in terms of the antisymmetric symbol $\epsilon^{\rho\sigma\alpha\beta}$ as $\varepsilon^{\rho\sigma\alpha\beta}=(1/\sqrt{-g})\epsilon^{\rho\sigma\alpha\beta}$, where $\epsilon^{0123}=1$. The third one is the dynamical term of the scalar field, with $\beta_0$ being another coupling parameter. As in Ref.\,\cite{Yagi2012pn}, we do not consider the potential of the scalar field. Finally, $\mathcal{L}_{m}$ is the Lagrangian density of the matter field.

The variation of the full action (\ref{action}) with respect to the metric $g^{\mu\nu}$ yields the modified field equation \cite{ChernSimons2003,Alexander2009},
\begin{equation}
\label{tensor-equation}
R_{\mu\nu}
-\frac{1}{2}g_{\mu\nu}R
+16\pi\alpha C_{\mu\nu}
=8\pi\left[T_{\mu\nu}^{(m)}+T_{\mu\nu}^{(\vartheta)}\right],
\end{equation}
where $R_{\mu\nu}$ is Ricci tensor and $C_{\mu\nu}$ is Cotton tensor defined as
\begin{equation}
\label{C-tensor}
C^{\mu\nu}=
-\varepsilon^{\rho(\mu|\alpha\beta|}
\left[\nabla_{\alpha}R^{\nu)}_{\ \beta}\right]
(\nabla_{\rho}\vartheta)
-\hat{R}^{\kappa(\mu|\rho|\nu)}
(\nabla_{\kappa}\nabla_{\rho}\vartheta).
\end{equation}
Note that the Cotton tensor $C_{\mu\nu}$ is traceless, $g^{\mu\nu}C_{\mu\nu}=0$, and satisfies the Bianchi identity, $\nabla^{\mu}C_{\mu\nu}=0$. $T_{\mu\nu}^{(m)}$ and $T_{\mu\nu}^{(\vartheta)}$ denote the energy-momentum tensors of the matter field and the DCS scalar field,
\begin{equation}
\label{scalar-EMT}
T_{\mu\nu}^{(\vartheta)}=\beta_0
\left[(\nabla_{\mu}\vartheta)(\nabla_{\nu}\vartheta)
-\frac{1}{2}g_{\mu\nu}
(\nabla_{\alpha}\vartheta)(\nabla^{\alpha}\vartheta)\right].
\end{equation}

The equation of the scalar field can also be derived by variation of the action (\ref{action}) to the scalar field $\vartheta$, which is
\begin{equation}
\label{scalar-equation}
\beta_0\Box^2_{g}\vartheta
=-\frac{\alpha}{4}R_{\nu\mu\alpha\beta}\hat{R}^{\mu\nu\alpha\beta}.
\end{equation}
We would like to mention here that when the coupling $\beta_0$ is $0$, the full action (\ref{action}) reduces to that of the non-dynamical Chern-Simons gravity. In this case, the scalar field equation (\ref{scalar-equation}) becomes an additional differential constraint, i.e., the {\em Pontryagin constraint} on the space of the allowed solutions, $R\hat{R}=0$. This work will not consider this case but only focus on the DCS gravity, in which the parameter $\beta_0\neq 0$.

This modification to GR leads to a series of parity-violating effects. One of the most important predictions is that the amplitude of the left-handed circular polarization mode of GWs increases (or decreases) during the propagation while the amplitude of the right-handed mode decreases (or increases). This phenomenon is always called amplitude birefringence of GWs \cite{Alexander2009,ZhaoLi2023,JinQiao2023}. The similar phenomenons (as well as the velocity birefringence) are investigated in other parity-violating theories, such as ghost-free parity-violating theory \cite{JinQiao2019,JinQiao2020}, Nieh-Yan gravity \cite{MingzheLi2020NY,MingzheLi2021NY,Chatzistavrakidis2022,Bombacigno2023}, Ho{\v{r}}ava-Lifshitz gravity \cite{Takahashi2009,TaoZhu2013,AnzhongWang2013}, parity-violating symmetric teleparallel gravity \cite{Conroy2019,MingzheLi2022,MingzheLi2022STG}, spatially covariant gravity \cite{XianGao2014, TaoZhu2022, TaoZhu2023}, and reviewed by Refs.\,\cite{WenZhao2020waveform,JinQiao2023,Jenks2023}. This effect greatly promotes the testing of parity symmetry in the gravitational sector by GW observation \cite{Maria2022,TaoZhu2022,Yunes2010PVtest,Mirshekari2012,WenZhao2020test,QiangWu2022,ChengGong2022,YifanWang2021,YifanWang2022,ZhichaoZhao2022}.

Due to the parity-violating property, the Pontryagin density generally disappears in spherically symmetric spacetime. For this reason, the Schwarzschild black hole is still an exact solution to DCS theory \cite{Grumiller2008}. The GWs radiated from binary Schwarzschild black holes are the same as those in GR unless the tidal deformation is considered \cite{Cardoso2017,Loutrel2022b}. In Refs.\,\cite{Yagi2012pn,Yagi2012gw,ZhaoLi2023} and this article, the spinning BBH is investigated. It should be noted that there is still no analytic rotating black hole solution with arbitrary spin in DCS theory. This work focuses on the slowly-rotating approximation \cite{Yunes2009,Yagi2012}. The DCS theory modifies the SS and MQ coupling \cite{Loutrel2018} between bodies and affects the gravitational waveforms. As explained in Ref.\,\cite{ZhaoLi2023}, since the time scale of the binary merger is much smaller than that of the cosmological expansion, the parity-violating effect of the DCS theory does not appear in the process of GW generation. And because the scalar field is coupled with the second-order metric perturbation and the Cotton tensor encoding the parity violation is traceless, none of the extra modes appears in this theory up to the leading order \cite{Wagle2019,ZhaoLi2023}, which is much different from other modified gravity. Unlike general scalar-tensor gravity, in DCS theory, the monopolar scalar charge of compact bodies is avoided \cite{Wagle2019b,Yunes2009,Yagi2012pn}, such that the spin-aligned binary black holes or binary neutron star systems will not produce scalar dipole radiation, but will produce quadrupole and higher-order radiation \cite{Yagi2012pn,ZhaoLi2023}. These scalar radiations still carry the energy and angular momentum, accelerating the orbital decay and binary coalescence.

\section{Eccentric Motion}
\label{sec:motion}
\subsection{\label{subsec:conserved-quantifies}Conserved Energy and Orbital Angular Momentum}

This work aims to extend the previous calculation \cite{Yagi2012pn,Yagi2012gw,ZhaoLi2023} in the quasi-circular case to the eccentric case. As a beginning, we first solve the EOM of the spin-aligned BBH system in DCS theory. For simplicity, throughout this article, only the Newtonian-order and DCS-modified terms are retained, and the PN correction of GR is dropped, as these contents can be found in previous publications, e.g., \cite{Blanchet2014} and references therein. 

We start from the conserved quantities of a BBH system, consisting of two spinning, well-separated bodies, whose mass and spin angular momentum vector are denoted by $m_A$ and $\bm{S}_{A}$, respectively. The DCS modified binding energy of such system in center-of-mass (COM) frame is given by Refs.\,\cite{Yagi2012gw,ZhaoLi2023}
\begin{equation}
\label{energy}
\varepsilon=\varepsilon_{\rm N}
+\delta\varepsilon.
\end{equation}
The Newtonian energy is
\begin{equation}
\label{0PN-energy}
\varepsilon_{\mathrm{N}}
=\frac{1}{2}v^2-\frac{m}{r}
\end{equation}
and its DCS modification is determined by so-called ``guess-work" \cite{Faye2006}, which gives
\begin{equation}
\label{DCS-energy}
\delta\varepsilon
=-\frac{1}{3}\delta\varpi
\left(\frac{m}{r}\right)^3,
\end{equation}
where $\bm{v}$ is the relative velocity of the BBH system, $m\equiv m_1+m_2$ is the total mass, and $r$ is the distance between two bodies. Following Ref.\,\cite{ZhaoLi2023}, the correction coefficient is defined as
\begin{equation}
\label{DCS-coefficient}
\delta\varpi\equiv
\zeta\left\{\frac{75}{256}\frac{1}{\nu}
\left(\frac{\bm{S}_1}{m_1^2}\cdot\frac{\bm{S}_2}{m_2^2}\right)
-\frac{603}{3584}\left[\frac{m^2}{m_1^2}\left(\frac{\bm{S}_1}{m_1^2}\right)^2
+\frac{m^2}{m_2^2}\left(\frac{\bm{S}_2}{m_2^2}\right)^2\right]\right\}.
\end{equation}
This coefficient shows two interactions between black holes induced by DCS modification, SS and MQ couplings. The symmetric mass ratio $\nu$ is defined by $\nu\equiv m_1m_2/m^2$ and the dimensionless coupling $\zeta$ is 
\begin{equation}
\zeta\equiv16\pi\frac{\alpha^2}{\beta_0m^4}.
\end{equation}
$\alpha$ and $\beta_0$ are the coupling parameters introduced in the DCS action (\ref{action}). As we have mentioned, the DCS theory only modifies the motion of the BBH system in quadratic-spin coupling due to parity violation, leaving the non-spin effects and SO coupling to be fully consistent with GR. It is noted that the terms with $\hat{\bm{n}}\cdot\bm{S}_A$ have been removed from Eq.\,(\ref{DCS-energy}) because of the spin-aligned assumption.

Another important conserved quantity, the OAM, in the COM frame is 
\begin{equation}
\label{angular-momentum}
\bm{h}=r(\hat{\bm{n}}\times\bm{v}).
\end{equation}
Unlike the conserved energy, however, the quadratic-spin effect does not modify the OAM up to 2PN approximation \cite{Will2014}. One can find the lowest-order, 3PN, correction in Refs.\,\cite{Bohe2015,Cho2022}. The conservation of OAM indicates that the motions of bodies are constrained on the orbital plane.

It is worth noting that these modifications to conserved quantities are valid in the following three approximations, small-coupling of $\sim\mathcal{O}(\zeta)$, slowly-rotating of $\sim\mathcal{O}(S^2)$, and 2PN of $\sim\mathcal{O}(v^4)$. Assuming the parameter $\alpha$, representing the strength of the interaction between scalar and tensor fields, to be weak and the bodies' spin to be sufficiently small admits an analytic solution to black hole spacetime \cite{Yunes2009,Yagi2012}, that is valid when dimensionless spin $\lesssim0.3$. The PN approximation allows an expansion in terms of the typical velocities of bodies \cite{Blanchet2014}. The quadratic-spin and 2PN approximation are equivalent to each other because the quadratic-spin effects first enter the 2PN correction \cite{Bohe2015,Cho2022}.

\subsection{\label{subsec:EOM}Equation of Motion}
The motions of the non-precessing BBH systems are constrained on the orbital plane. It is convenient to describe the motion using polar coordinate, radial coordinate $r$, and azimuth coordinates $\phi$. Thus, one can define the relative direction vector, $\hat{\bm{n}}=(\cos\phi, \sin\phi, 0)$, pointing $2$-nd body from $1$-st body and $\hat{\bm{\lambda}}=(-\sin\phi, \cos\phi, 0)$ as another orthogonal direction on the orbital plane. Therefore, the relative velocity $\bm{v}$ can be expanded as $\bm{v}=\dot{r}\hat{\bm{n}}+r\dot{\phi}\hat{\bm{\lambda}}$, where ``dot" means the derivative with respect to the time $t$. Combining the above definitions and conserved quantifies (\ref{energy}, \ref{angular-momentum}), one can re-derive the EOM of the BBH system. The radial and azimuth equations are  
\begin{equation}
\label{radial-equation}
\dot{r}^2=2\varepsilon+2\gamma-j^2\gamma^2
+\frac{2}{3}\delta\varpi\gamma^3,\quad\text{and}\quad
m\dot{\phi}=j\gamma^2,
\end{equation}
respectively, where $\gamma\equiv m/r$ and $mj\equiv h=|\bm{h}|$. The radial equation can be solved through the following integration, 
\begin{equation}
\label{time-integral}
t-t_0=\pm\frac{m}{j}
\int\frac{(1
+\frac{2}{3}\frac{\delta\varpi}{j^4})
+\frac{1}{3}\frac{\delta\varpi}{j^2}\gamma}
{\gamma^2\sqrt{(\gamma_+-\gamma)
(\gamma-\gamma_-)}}\dd\gamma
=\pm\int T(\gamma)\dd\gamma.
\end{equation}
In Eq.\,(\ref{time-integral}), $\gamma_{\pm}$ are defined as the periastron and apastron of the binary system, which is given perturbatively by
\begin{equation}
\label{periastron-apastron}
\gamma_{\pm}=\frac{1\pm\sqrt{1+2j^2\varepsilon}}{j^2}
\left[1\pm\frac{1}{3}\frac{\delta\varpi}{j^4}
\frac{(1\pm\sqrt{1+2j^2\varepsilon})^2}{\sqrt{1+2j^2\varepsilon}}\right].
\end{equation}
$t_0$ is an integration constant, representing the time when the bodies first pass through the periastron. 

Similarly, the solution to the azimuth equation has the following form,
\begin{equation}
\label{azimuth-integral}
\phi-\phi_0
=\pm\int\frac{
(1+\frac{2}{3}\frac{\delta\varpi}{j^4})
+\frac{1}{3}\frac{\delta\varpi}{j^2}\gamma}
{\sqrt{(\gamma_+-\gamma)
(\gamma-\gamma_-)}}\dd\gamma
=\pm\int\Phi(\gamma)\dd\gamma,
\end{equation}
where $\phi_0$ is another integration constant, representing the initial azimuth coordinate of the periastron. 

Finally, the extra signs in the integrations (\ref{time-integral}, \ref{azimuth-integral}) are determined by whether the bodies move from the periastron to the apastron or vice versa. When the black hole moves from the periastron to the apastron, these integrations are evaluated as
\begin{equation}
\label{sign-periaston-to-apastron}
t-t_0=\int_{\gamma}^{\gamma_+}T(\gamma)\dd\gamma,\quad\text{and}\quad
\phi-\phi_0=\int_{\gamma}^{\gamma_+}\Phi(\gamma)\dd\gamma,
\end{equation}
and, inversely, they are evaluated as
\begin{equation}
\label{sign-apastron-periaston}
t-t_0=\int_{\gamma_{-}}^{\gamma}T(\gamma)\dd\gamma
+\int_{\gamma_{-}}^{\gamma_{+}}T(\gamma)\dd\gamma,\quad\text{and}\quad
\phi-\phi_0=\int_{\gamma_{-}}^{\gamma}\Phi(\gamma)\dd\gamma
+\int_{\gamma_{-}}^{\gamma_{+}}\Phi(\gamma)\dd\gamma.
\end{equation}

We would like to note that, when the orbits are no longer closed, the concept of the period is obscured. In general, a time period is considered as the time interval between two consecutive passes of a body through the periastron, and the azimuth interval that the bodies pass during this period is considered an azimuth period. In the Newtonian limit, these two periods are related by Kepler's third law. However, after considering the PN correction, the situation becomes different. It is convenient for us to introduce an alternative angular variable, true anomaly \cite{Roy2020} denoted by $V$, which is defined as the difference between the azimuth coordinate and the current periastron. The true anomaly passes through $2\pi$ during a period, while the azimuth is more advanced. This means that the eccentric motion presents a {\em doubly periodic structure} in the PN framework \cite{Blanchet2014}.

\subsection{Quasi-Keplerian Parameterization}
The above integrations (\ref{time-integral}, \ref{azimuth-integral}) in the limit ($\delta\varpi=0$) give an elliptic orbit described by two unique parameters, the eccentricity $e$ and the semimajor axis $a$. The trajectory, $r=r(\phi)$, of relative motion is $r=a(1-e^2)/[1-e\cos(\phi-\phi_0)]$. However, when considering the PN corrections, the gravitational interaction violates the inverse-square law. The eccentricity and semimajor axis of an unclosed orbit cannot be well defined. To deal with that, the corresponding QK parameterization introduces two new elements, the ``radial" eccentricity $e_r$ and the ``radial" semimajor axis $a_r$ (eccentricity and semimajor axis in short), defined through periastron and apastron by
\begin{equation}
\label{elements-definition}
a_r\equiv2m\frac{\gamma_+\gamma_-}{\gamma_++\gamma_-},\quad\text{and}\quad
e_r\equiv\frac{\gamma_+-\gamma_-}{\gamma_++\gamma_-},
\end{equation}
respectively, to simplify integrations (\ref{time-integral}, \ref{azimuth-integral}). The modified trajectory is parameterized through
\begin{equation}
\label{parameterization}
\gamma=\frac{\xi}{1-e_r\cos u},\quad\text{and}\quad
\xi=\frac{m}{a_{r}}.
\end{equation}
$u$ is the eccentric anomaly related to the true anomaly by some geometric relationships,
\begin{equation}
\label{relation-true-eccentric-anomaly}
\cos V=\frac{\cos u-e_r}{1-e_r\cos u},\quad\text{and}\quad
\sin V=\frac{\sqrt{1-e_r^2}\sin u}{1-e_r\cos u},
\end{equation}
or equivalently, the direct definition is
\begin{equation}
\label{true-anomaly-definition}
V\equiv2\arctan\left[\sqrt{\frac{1+e_r}{1-e_r}}
\tan\left(\frac{u}{2}\right)\right].
\end{equation}
When setting the eccentricity, $e_r$, to zero, the relative distance $r$ reduces to a constant, such that the quasi-elliptic orbits become quasi-circular ones.

Using Eq.\,(\ref{periastron-apastron}), one can re-express the elements (\ref{elements-definition}) in terms of the energy and OAM,
\begin{equation}
\label{elements-conserved-quantities}
\xi=-2\varepsilon\left(1-\frac{2}{3}\delta\varpi\frac{\varepsilon}{j^2}\right),\quad\text{and}\quad
e_{r}=\sqrt{1+2j^2\varepsilon}
\left(1-\frac{4}{3}\delta\varpi
\frac{\varepsilon}{j^2}
\frac{1+j^2\varepsilon}{1+2j^2\varepsilon}\right).
\end{equation}
We note that neither of the above elements is a geometric quantity like that in the Newtonian case, but just two parameters related to the conserved quantities, $\varepsilon$ and $j$. Inversely, the conserved energy and OAM also can be represented by these elements, they are
\begin{equation}
\label{conserved-quantities-elements}
\varepsilon=-\frac{\xi}{2}\left(
1-\frac{1}{3}\delta\varpi\frac{\xi^2}{1-e_r^2}\right),\quad\text{and}\quad
j=\frac{\sqrt{1-e_r^2}}{\sqrt{\xi}}\left[
1+\frac{1}{6}\delta\varpi
\left(\frac{\xi}{1-e_r^2}\right)^2(3+e_r^2)\right],
\end{equation}
respectively. The parameterized time integration (\ref{time-integral}) and azimuth integration (\ref{azimuth-integral}) are given by
\begin{equation}
\label{parameterized-integral}
\begin{aligned}
t&=\frac{m}{j}\frac{\sqrt{1-e_r^2}}{\xi^2}
\int_0^u\left[-\left(1+
\frac{2}{3}\frac{\delta\varpi}{j^4}\right)
(1-e_r\cos u)
-\frac{1}{3}\frac{\delta\varpi}{j^2}\xi\right]\dd u,\\
\text{and}\quad\phi&=\sqrt{1-e_r^2}
\int_0^u\frac{1}{(1-e_r\cos u)^2}
\left[-\left(1
+\frac{2}{3}\frac{\delta\varpi}{j^4}\right)
(1-e_r\cos u)
-\frac{1}{3}\frac{\delta\varpi}{j^2}\xi\right]\dd u.
\end{aligned}
\end{equation}
Without the loss of generality, the integration constants, $t_0$ and $\phi_0$, have been taken as zero.

\subsection{Solution and Time, Azimuth Period}
After parameterization, the integrations (\ref{parameterized-integral}) can be evaluated directly through integral formulas
\begin{equation}
\label{integral-formula}
\int\frac{\dd u}{(1-e_r\cos u)^2}
=\frac{1}{(1-e_r^2)^{3/2}}(e_r\sin V+V),
\quad\text{and}\quad
\int\frac{\cos u\dd u}{(1-e_r\cos u)^2}
=\frac{1}{(1-e_r^2)^{3/2}}(\sin V+e_rV).
\end{equation}
Starting from Eqs.\,(\ref{parameterized-integral}) and (\ref{integral-formula}), we get
\begin{equation}
\label{time-azimuth-integral-results}
\begin{aligned}
t(u)&=\frac{m}{j}\frac{\sqrt{1-e_r^2}}{\xi^2}
\left\{\left[\left(1+
\frac{2}{3}\frac{\delta\varpi}{j^4}\right)
+\frac{1}{3}\frac{\delta\varpi}{j^2}\xi\right]u
-\left(1+
\frac{2}{3}\frac{\delta\varpi}{j^4}\right)e_r\sin u\right\},\\
\text{and}\quad\phi(u)&=\left[
\left(1+
\frac{2}{3}\frac{\delta\varpi}{j^4}\right)
+\frac{1}{3}\frac{\delta\varpi}{j^2}
\left(\frac{\xi}{1-e_r^2}\right)\right]V
+e_r\left[\frac{1}{3}\frac{\delta\varpi}{j^2}
\left(\frac{\xi}{1-e_r^2}\right)\right]\sin V.
\end{aligned}
\end{equation}
From the integration results (\ref{time-azimuth-integral-results}),  we can conclude two important parameters, the time period and azimuth period mentioned in \ref{subsec:EOM}, are given by
\begin{equation}
T\equiv t(u=2\pi)
=2\pi m(-2\varepsilon)^{-3/2},\quad
\text{and}\quad K\equiv\phi(u=2\pi)
=2\pi\left(1+\frac{\delta\varpi}{j^4}\right),
\end{equation}
respectively. In the Newtonian order, the orbit of a binary system is standard ellipses. Such that the black holes can accurately return to their periastron, completing an azimuth period, within a time period. However, in the higher-PN order approximation, the azimuthal motion of BBH exceeds $2\pi$ within a time period. The residual azimuth motion is the well-known periastron-advance effect \cite{Damour1985,Damour1988,Wex1995,Memmesheimer2004,Will2014}.

\subsection{Final Parameterization}
In this subsection, we summarize the final results of quasi-Keplerian motion. The time and azimuth of the bodies are shown as the function of the true anomaly or eccentric anomaly (\ref{time-azimuth-integral-results}), which are
\begin{equation}
\label{time-results}
\frac{2\pi}{T}t(u)
=u-e_t\sin u,
\end{equation}
and
\begin{equation}
\label{azimuth-results}
\frac{2\pi}{K}\phi(u)
=2\arctan\left[\sqrt{\frac{1+e_{\phi}}{1-e_\phi}}\tan\left(\frac{u}{2}\right)\right]\equiv v,
\end{equation}
respectively. Eq.\,(\ref{time-results}) is usually called \emph{modified Keplerian equation}. Together with $r=a_r(1-e_r\cos u)$, we complete the all steps for parameterization. In the final results (\ref{time-results}, \ref{azimuth-results}), introducing the ``time" and ``azimuth" eccentricities, $e_t$ and $e_\phi$, aims to simplify the complicated expressions. They are not independent elements but also depend on the conserved quantities by
\begin{equation}
e_t=e_r\left(1
+\frac{2}{3}\delta\varpi\cdot
\frac{\varepsilon}{j^2}\right),\quad\text{and}\quad
e_\phi=e_r\left(1-\frac{2}{3}\delta\varpi\cdot
\frac{\varepsilon}{j^2}\right).
\end{equation}
At the same time, we  can also write them in terms of ``radial" elements by
\begin{equation}
\label{et-ephi-er}
e_{t}=e_{r}\left(1
-\frac{1}{3}\delta\varpi\cdot
\frac{\xi^2}{1-e_r^2}\right),\quad\text{and}\quad
e_{\phi}=e_{r}\left(1
+\frac{1}{3}\delta\varpi\cdot
\frac{\xi^2}{1-e_r^2}\right).
\end{equation}
Additionally, the time and azimuth periods are also expressed by elements $a_r$ and $e_r$ as follows,
\begin{equation}
\label{T-K}
T=2\pi\frac{m}{\xi^{3/2}}\left(
1+\frac{1}{2}\delta\varpi\cdot
\frac{\xi^2}{1-e_r^2}\right),\quad\text{and}\quad
K=2\pi\left[1+\delta\varpi\left(\frac{\xi}{1-e_r^2}\right)^2\right].
\end{equation}
$\xi$ is one of the bookkeepers of the PN order. An $n$-PN term generally contains a factor $\xi^{n}$. And we can find again the DCS theory modifies the BBH motion at 2PN approximation.

The portion of the azimuth period $K$ that exceeds $2\pi$ represents the periastron-advance effect, and the precession rate is defined as 
\begin{equation}
\beta\equiv\frac{K}{2\pi}-1=\delta\varpi\cdot\frac{\xi^2}{(1-e_r^2)^2}.
\end{equation}
After a periodic motion, the periastron advances an extra angle $2\pi\beta$ to the last period. The well-known 1PN analogue is $6\pi\xi/(1-e_r^2)$, successfully explaining the perihelion advance of Mercury \cite{Will2014}. One can find that the DCS modification enters 2PN order, as we have shown in previous work \cite{ZhaoLi2023}. Another important point is that the precession rate of the periastron is nonvanishing when eccentricity is zero. In other words, the precession effect still exists even in circular-orbit motion, which is incomprehensible. This anomaly originates from the ill definition of the periastron of the circular orbit. This anomaly can be eliminated in subsequent derivation by defining the gauge-invariant parameter, orbital frequency
\begin{equation}
\Omega\equiv\frac{K}{T}=\frac{\xi^{3/2}}{m}
\left[1+\frac{\delta\varpi}{2}\frac{\xi^2}{(1-e_r^2)^2}(1+e_r^2)\right]
\end{equation}
or the dimensionless frequency
\begin{equation}
    x\equiv(m\Omega)^{2/3}
\end{equation}
which is another bookkeeper of the PN order. $n$-PN terms generally carry factors of $x^{n}$. Additionally, Kepler's third law with higher-order modification, relating the elements and the frequency, is
\begin{equation}
\xi=x\left[1-\frac{\delta\varpi}{3}
\frac{(1+e_r^2)}{(1-e_r^2)^2}x^2\right].
\end{equation}

For the  calculation in the next section, we also present the parameterization of the time derivatives of radial and azimuth coordinates, $\dot{r}$ and $\dot{\phi}$, 
\begin{equation}
\label{dotr-dotphi}
\begin{aligned}
\dot{r}&
=\sqrt{\xi}\frac{e_r\sin u}{1-e_r\cos u}
\left[1-\frac{1}{6}\delta\varpi\cdot
\frac{\xi^2(3-e_r\cos u)}{(1-e_r^2)(1-e_r\cos u)}
\right],\\
\text{and}\quad\dot{\phi}&
=\frac{\xi^{3/2}}{m}\frac{\sqrt{1-e_r^2}}
{(1-e_r\cos u)^2}
\left[1+\frac{1}{6}\delta\varpi\cdot
\frac{\xi^2(3+e_r^2)}{(1-e_r^2)^2}\right].
\end{aligned}
\end{equation}

\section{\label{sec:waveform}Gravitational Radiation}
At present, we have provided a complete solution to BBH motion in DCS gravity, Eqs.\,(\ref{time-results}) and (\ref{azimuth-results}), equipped by parameterization, $r=a_r(1-e_r\cos u)$. Now, we turn to consider the radiation field observed at the far zone with inclination angle $\iota$ and azimuth angle $\omega$. The line-of-sight vector is $\hat{\mathbf{N}}=(\sin\iota\sin\omega,\sin\iota\cos\omega,\cos\iota)$ and distance is $R$. As we have mentioned above, due to the periastron advance, the azimuth coordinate $\phi$ is not considered to be a suitable periodic angular variable. In contrast, the true anomaly $V$ and mean anomaly $\ell$ are usually used to describe the waveform. In this section, we focus on the true-anomaly representation. From Eq.\,(\ref{azimuth-results}), the relationship between $\phi$ and $V$ is given by
\begin{equation}
\label{phi-V}
\phi=2K\arctan\left[\sqrt{\frac{1+e_{\phi}}{1-e_\phi}}\tan\left(\frac{u}{2}\right)\right]
\simeq V(1+\beta)+\frac{1}{3}\delta\varpi\cdot\frac{e_r\xi ^2}{(1-e_r^2)^{2}}\sin V,
\end{equation}
up to the linear order of coupling $\zeta$. This relation gives the periodic function, 
\begin{equation}
\label{sin-cos-phi-V}
\begin{aligned}
\sin\phi
&=\sin[(1+\beta)V]
+\frac{1}{3}\delta\varpi\cdot\frac{e_r\xi ^2}{(1-e_r^2)^{2}}
\cos[(1+\beta)V]\sin V,\\
\text{and}\quad\cos\phi
&=\cos[(1+\beta)V]
-\frac{1}{3}\delta\varpi\cdot\frac{e_r\xi ^2}{(1-e_r^2)^{2}}\sin[(1+\beta)V]\sin V.
\end{aligned}
\end{equation}
Although the parameter $\beta$ is always a perturbative quantity of order $\mathcal{O}(\zeta\xi^2)$, $\beta V$ cannot be regarded as a small one as $V$ monotonically increases over time. In comparison, the functions $\sin V$ and $\cos V$, with upper and lower limits, always allow a Taylor expansion in terms of $\delta\varpi\sin V$ and $\delta\varpi\cos V$.

Equation.\,(\ref{sin-cos-phi-V}) presents a doubly-periodic structure in the BBH motion. The first period is given by $\sin(1+\beta)V$ and $\cos(1+\beta)V$. The azimuth of bodies passes through a $2\pi$ angle within this period. The second one is given by $\sin V$ and $\cos V$. The bodies return to the periastron within this period. This structure will enter the gravitational waveform that will be displayed soon.

\subsection{Scalar Radiation}
The full scalar radiation at infinity is given by our previous work \cite{ZhaoLi2023}, 
\begin{equation}
\vartheta
=\frac{2m\nu}{R}\cdot
\frac{5}{16}\gamma^2
\frac{\alpha}{\beta_0m^2}
\frac{1}{\nu}
[(\hat{\bm{n}}\cdot\tilde{\bm{\Delta}})
+(\hat{\mathbf{N}}\cdot\tilde{\bm{\Delta}})
(\hat{\mathbf{N}}\cdot\hat{\bm{n}})].
\end{equation}
The difference between bodies' spins is 
\begin{equation}
\tilde{\bm{\Delta}}\equiv
\frac{m_{2}}{m}\frac{\bm{S}_{1}}{m_1^2}
-\frac{m_{1}}{m}\frac{\bm{S}_{2}}{m_2^2},
\end{equation}
producing a minus sign when exchanging labels $1\leftrightarrow2$, and vanishing when the masses and spins are exactly equal for these two black holes, $m_1=m_2$ and $\bm{S}_1=\bm{S}_2$, meaning again that scalar field is a pseudoscalar field. For non-precessing binaries, in which $(\hat{\bm{n}}\cdot\tilde{\bm{\Delta}})=0$, the radiation field reduces to 
\begin{equation}
\label{vartheta-form}
\vartheta=\frac{2m\nu}{R}\cdot
\frac{5}{16}\frac{\alpha}{\beta_0m^2}
\frac{\gamma^2}{\nu}
(\hat{\mathbf{N}}\cdot\tilde{\bm{\Delta}})
(\hat{\mathbf{N}}\cdot\hat{\bm{n}}).
\end{equation}
Substituting Eq.\,(\ref{sin-cos-phi-V}) into (\ref{vartheta-form}), we get
\begin{equation}
\label{vartheta-result}
\vartheta
=-\frac{2m\nu}{R}\frac{5}{16}
\frac{\alpha}{\beta_0m^2}
\frac{\xi^2}{\nu}
\frac{(1+e_r\cos V)^2}{(1-e_r^2)^2}
\tilde{\Delta}\sin\iota\cos\iota
\sin(V+\omega),
\end{equation}
with $\tilde{\Delta}\equiv|\tilde{\bm{\Delta}}|$. When setting the eccentricity and observation azimuth to zero, this expression becomes that in the quasi-circular case \cite{Yagi2012pn,ZhaoLi2023}.

\subsection{Tensor Radiation}
In the PN framework, the gravitational waveform contains an ``instantaneous" term, depending on the state of the binary at the retarded time only, and a ``tail" term, which is sensitive to the wave field at all previous time \cite{Blanchet1995}. The quadratic-spin correction does not change the tail term in a non-precessing system \cite{Bohe2015,ZhaoLi2023}, such that we focus on the ``instantaneous" term only. In the transverse-traceless (TT) gauge, the metric tensor of GW is
\begin{equation}
\label{inst-waveform}
(h_{ij}^{\mathrm{TT}})_{\mathrm{inst}}
=\frac{2\nu m}{R}
\xi_{ij}^{\mathrm{TT}}
=\frac{2\nu m}{R}
\hat{\Lambda}_{ij,kl}\xi_{kl},
\end{equation}
where $\hat{\Lambda}_{ij,kl}$ is the TT-projection operator, defined as $\hat{\Lambda}_{ij,kl}(\hat{\mathbf{N}})\equiv\Pi_{ik}\Pi_{jl}-(1/2)\Pi_{ij}\Pi_{kl}$, with $\Pi_{ij}\equiv\delta_{ij}-\hat{N}_i\hat{N}_{j}$. The reduced metric tensor is decomposed into a Newtonian term and the DCS modification, 
\begin{equation}
\label{total-metric}
\xi_{ij}=\xi_{ij}^{(0)}
+\delta\xi_{ij}.
\end{equation}
The Newtonian term is 
\begin{equation}
\label{Newtonian-metric}
\xi_{ij}^{(0)}=2(v^{i}v^j-\gamma\hat{n}^{i}\hat{n}^j),
\end{equation}
and its components are
\begin{equation}
\label{Newtonian-metric-components}
\begin{aligned}
\xi_{11}^{(0)}&=2
\left[(\dot{r}\cos\phi
-r\dot{\phi}\sin\phi)^2
-\frac{m}{r}\cos^2\phi\right],\\
\xi_{12}^{(0)}&=2
\left[(\dot{r}\sin\phi
+r\dot{\phi}\cos\phi)
(\dot{r}\cos\phi
-r\dot{\phi}\sin\phi)
-\frac{m}{r}\sin\phi\cos\phi\right],\\
\xi_{22}^{(0)}&=2
\left[(\dot{r}\sin\phi+r\dot{\phi}\cos\phi)^2
-\frac{m}{r}\sin^2\phi\right].
\end{aligned}
\end{equation}
The DCS modification is \cite{ZhaoLi2023}
\begin{equation}
\label{DCS-metric}
\delta\xi_{ij}
=-2\cdot\delta\varpi
\cdot\gamma^3\hat{n}_i\hat{n}_j
=-\delta\varpi
\left(\frac{m}{r}\right)^3
\left(
\begin{array}{cc}
2\cos^2\phi
&2\sin\phi\cos\phi\\
2\sin\phi\cos\phi
&2\sin^2\phi
\end{array}
\right).
\end{equation}

Using the rotation matrix,
\begin{equation}
\label{rotation-matrix}
\bm{\mathcal{R}}\equiv\bm{\mathcal{R}}_{z}(\omega)\bm{\mathcal{R}}_{x}(\iota)
=\left(\begin{array}{ccc}
\cos\omega&-\sin\omega & 0 \\
\sin\omega& \cos\omega & 0 \\
0 & 0 & 1 \\
\end{array}\right)
\left(\begin{array}{ccc}
1 & 0 & 0 \\
0 & \cos\iota& -\sin\iota\\
0 & \sin\iota & \cos\iota\\
\end{array}
\right),
\end{equation}
we transform the metric tensor from the binary frame to the propagation frame along the observational direction, $\xi_{ij}(\hat{\mathbf{N}})=\mathcal{R}_{ik}\mathcal{R}_{jl}\cdot\xi_{kl}$. TT projecting, $\hat{\Lambda}_{ij,kl}\xi_{kl}$, gives the waveforms $\xi_{ij}^{\mathrm{TT}}$ in TT gauge and the plus mode and the cross mode are just 
\begin{equation}
    \xi_{+}\equiv\xi_{11}^{\mathrm{TT}} = \xi_{+}^{(0)} + \delta\xi_{+} \quad \text{and} \quad \xi_{\times}\equiv\xi_{12}^{\mathrm{TT}}= \xi_{\times}^{(0)}+ \delta\xi_{\times},
\end{equation}
where
\begin{equation}
\label{Newtonian-polarization-in-rphi}
\begin{aligned}
\xi_{+}^{(0)}&=
\left[\left(\dot{r}^2
-r^2\dot{\phi}^2
-\gamma\right)\cos(2\omega)
-2r\dot{r}\dot{\phi}
\sin(2\omega)\right]
\frac{1+\cos^2\iota}{2}\cos(2\phi)\\
&+\left[-\left(\dot{r}^2
-r^2\dot{\phi}^2-\gamma\right)
\sin(2\omega)
-2r\dot{r}\dot{\phi}
\cos(2\omega)\right]
\frac{1+\cos^2\iota}{2}\sin(2\phi)
+\frac{1}{2}\sin^2\iota
\left[(\dot{r}^2+r^2\dot{\phi}^2)
-\gamma\right],\\
\xi_{\times}^{(0)}
&=\left[\sin(2\omega) \left(\dot{r}^2-r^2\dot{\phi}^2-\gamma\right)
+2r\dot{r}\dot{\phi}
\cos(2\omega)\right]
\cos\iota\cos(2\phi)\\
&+\left[\cos(2\omega)\left(\dot{r}^2
-r^2\dot{\phi}^2-\gamma\right)
-2r\dot{r}\dot{\phi}
\sin(2\omega)\right]
\cos\iota\sin(2\phi),
\end{aligned}
\end{equation}
and
\begin{equation}
\label{DCS-polarization-in-rphi}
\begin{aligned}
\delta\xi_{+}
&=-\delta\varpi\cdot\gamma^3
\left[\frac{1+\cos^2\iota}{2}
+\frac{\sin^2\iota}{2}
\cos(2\phi+2\omega)\right],\\
\delta\xi_{\times}
&=-\delta\varpi\cdot\gamma^3
\cos\iota\sin(2\phi+2\omega).
\end{aligned}
\end{equation}
As we have commented at Sec \ref{sec:DCS}, there are no extra modes in the gravitational radiation. 

\subsection{Waveforms in Terms of True Anomaly}
In this subsection, we present the final results of the GW polarizations in terms of the true anomaly. Substituting the $r=a_r(1-e_r\cos u)$, and Eqs.\,(\ref{sin-cos-phi-V}, \ref{dotr-dotphi}, \ref{relation-true-eccentric-anomaly}) into (\ref{Newtonian-polarization-in-rphi}, \ref{DCS-polarization-in-rphi}), the final waveforms are expressed as
\begin{equation}
\label{Newtonian-polarization-in-V}
\begin{aligned}
\xi_{+}^{(0)}
&=\frac{\xi}{1-e_r^2}
\left\{-\frac{1}{2}e_r^2\Big[(1+\cos^2\iota)
\cos(2\beta V+2\omega)
-\sin^2\iota\Big]\right.\\
&\qquad-\frac{5}{4}e_r(1+\cos^2\iota)
\cos[(1+2\beta)V+2\omega]
+\frac{1}{2}e_r\sin^2\iota\cos V\\
&\left.\qquad\qquad-(1+\cos^2\iota)
\cos[(2+2\beta)V+2\omega]
-\frac{1}{4}e_r(1+\cos^2\iota)
\cos[(3+2\beta)V+2\omega]\right\},\\
\xi_{\times}^{(0)}
&=\frac{\xi}{1-e_r^2}
\left\{-e_r^2\sin(2\beta V+2\omega)
-\frac{5}{2}e_r\sin[(1+2\beta) V+2\omega]\right.\\
&\left.\qquad-2\sin[(2+2\beta)V+2\omega]
-\frac{1}{2}e_r\sin[(3+2\beta) V+2\omega]\right\}\cos\iota,
\end{aligned}
\end{equation}
and
\begin{equation}
\label{DCS-polarization-in-V}
\begin{aligned}
\delta\xi_{+}&=
\frac{\xi^3}{(1-e_r^2)^3}
\delta\varpi\left\{
-\frac{1}{12}e_r(11+6e_r^2)(1+\cos^2\iota)
\cos[(1+2\beta)V+2\omega]\right.\\
&-\frac{1}{8}e_r(4+e_r^2)\sin^2\iota\cos V
+\frac{3}{8}e_r^3
(1+\cos^2\iota)\sin(2\beta V+2\omega)\sin V\\
&-\frac{1}{4}e_r^2\sin^2\iota\cos 2V
-\frac{1}{4}(4+7e_r^2)(1+\cos^2\iota)
\cos[(2+2\beta)V+2\omega]\\
&-\frac{1}{24}e_r^3\sin^2\iota\cos3V
-\frac{1}{48}e_r(76+13e_r^2)(1+\cos^2\iota)
\cos[(3+2\beta)V+2\omega]\\
&\left.-\frac{13}{24}e_r^2
(1+\cos^2\iota)\cos[(4+2\beta)V+2\omega]
-\frac{1}{16}e_r^3
(1+\cos^2\iota)\cos[(5+2\beta)V+2\omega]
\right\},\\
\delta\xi_{\times}&=
\frac{\xi^3}{(1-e_r^2)^3}
\delta\varpi\left\{
-\frac{1}{6}e_r(11+6e_r^2)
\cos\iota\sin[(1+2\beta)V+2\omega]
-\frac{3}{4}e_r^3
\cos\iota\cos(2\beta V+2\omega)\sin V\right.\\
&-\frac{1}{2}(4+7e_r^2)
\cos\iota\sin[(2+2\beta)V+2\omega]
-\frac{1}{24}e_r(76+13e_r^2)
\cos\iota\sin[(3+2\beta)V+2\omega]\\
&\left.-\frac{13}{12}e_r^2
\cos\iota\sin[(4+2\beta)V+2\omega]
-\frac{1}{8}e_r^3
\cos\iota\sin[(5+2\beta)V+2\omega]\right\}.
\end{aligned}
\end{equation}
The waveforms for quasi-circular case with $e_r=0$ and $\omega=0$ are
\begin{equation}
\label{Newtonian-polarization-circular}
\begin{aligned}
\xi_{+}^{(0)}
&=-\xi(1+\cos^2\iota)
\cos(2+2\beta)V
=-\xi(1+\cos^2\iota)\cos2\phi,\\
\xi_{\times}^{(0)}
&=-2\xi\cos\iota\sin(2+2\beta)V
=-2\xi\cos\iota\sin2\phi,
\end{aligned}
\end{equation}
and
\begin{equation}
\label{DCS-polarization-circular}
\begin{aligned}
\delta\xi_{+}&=-\delta\varpi\cdot\xi^3
(1+\cos^2\iota)\cos(2+2\beta)V
=-\delta\varpi\cdot\xi^3
(1+\cos^2\iota)\cos2\phi\\
\delta\xi_{\times}&=-2\cdot\delta\varpi\cdot\xi^3
\cos\iota\sin(2+2\beta)V
=-2\cdot\delta\varpi\cdot\xi^3
\cos\iota\sin2\phi,
\end{aligned}
\end{equation}
returning to that predicted by our previous work \cite{ZhaoLi2023}.

We briefly summarize the features of these waveforms. Firstly and most importantly, there are no extra polarization modes, the conclusion of Ref.\,\cite{Wagle2019,ZhaoLi2023}. The scalar field does not produce the breathing mode as in massless Brans-Dicke gravity \cite{XingZhang2017BD} and longitudinal mode as in massive Brans-Dicke gravity \cite{TanLiu2020BD}. Secondly, there are no parity-violating effects, because of the neglect of cosmic expansion. Thirdly, there is a more complicated frequency spectrum compared with the quasi-circular case. The GWs are emitted at a set of discrete phases $\{V,2V,3V\}$ at Newtonian order and $\{V,2V,3V,4V,5V\}$ at DCS order, rather than at a single phase $2V$ for circular motion. These extra frequency modes will disappear at the circular limit. Finally, the waveforms are modulated by the periastron-advance effect as we have discussed when calculating the BBH motion. This modulation is non-vanishing for $e_r=0$ because the periastron is ill-defined in the circular limits. To complete the calculation in Eqs.\,(\ref{Newtonian-polarization-circular}) and (\ref{DCS-polarization-circular}). one needs to take the substitution $(1+\beta)V\rightarrow\phi$.

\section{Radiation Reaction}
\label{sec:radiation-reaction}
\subsection{Energy Flux}
Although the scalar radiation does not influence the GW polarizations, it carries the energy and angular momentum, changing the orbital elements of BBH orbit. The total radiated energy carried by the scalar radiation is defined as
\begin{equation}
\label{scalar-energy-flux-definition}
\mathcal{F}_{S}
=\beta_0R^2\oint_{\partial\Omega}
\langle\dot{\vartheta}^2\rangle d\Omega,
\end{equation}
in which the orbital average is defined as
\begin{equation}
\label{average-definition}
\langle\dot{\vartheta}^2\rangle
=\frac{1}{T}\int_0^{T}
\left(\frac{\partial\vartheta}{\partial t}\right)^2dt
=\frac{1}{T}\int_0^{2\pi}
\left(\frac{\partial\vartheta}
{\partial V}\right)^2
\frac{dV}{du}
\frac{du}{dt}dV.
\end{equation}
Using the definition of true anomaly (\ref{relation-true-eccentric-anomaly}) and the time integration (\ref{time-integral}), we obtain
\begin{equation}
\frac{dV}{du}
=\frac{\sqrt{1-e_r^2}}
{1-e_r\cos u}
=\frac{1+e_r\cos V}
{\sqrt{1-e_r^2}}
\end{equation}
and 
\begin{equation}
\frac{dt}{du}\approx
\frac{m}{\xi^{3/2}}
\frac{1-e_r^2}{1+e_r\cos V}
\left[1+\frac{\delta\varpi}{6}
\frac{\xi^2}{(1-e_r^2)^2}
(3+2e_r\cos V-e_r^2)\right]
\end{equation}
up to the linear order of the coupling. After a long calculation, the orbital average (\ref{average-definition}) is
\begin{equation}
\beta_0\langle\dot{\vartheta}^2\rangle
=\frac{1}{8\pi}
\frac{25}{256}
\frac{\nu^2\xi^7}{(1-e_{r}^2)^{11/2}}\Delta^2\sin^2\iota\cos^2\iota
\left[1+\frac{19}{2}e_r^2
+\frac{69}{8}e_r^4
+\frac{9}{16}e_r^6
-\frac{1}{4}e_{r}^2
\left(1+5e_r^2
+\frac{9}{16}e_r^4\right)
\cos2\omega\right],
\end{equation}
where
\begin{equation}
\label{delta2}
\Delta^2\equiv(\zeta/\nu^2)\tilde{\Delta}^2=\zeta
\left\{-\frac{2}{\nu}\left(\frac{\bm{S}_{1}}{m_1^2}\cdot\frac{\bm{S}_{2}}{m_2^2}\right)+\left[\frac{m^2}{m_1^2}\left(\frac{\bm{S}_{1}}{m_1^2}\right)^2+\frac{m^2}{m_2^2}\left(\frac{\bm{S}_{2}}{m_2^2}\right)^2\right]\right\}.
\end{equation}
This new symbol has the same structure of $\delta\varpi$ (\ref{DCS-coefficient}) and also shows the SS and MQ coupling modified by DCS theory. The solid angle integration in Eq.\,(\ref{scalar-energy-flux-definition}) is 
\begin{equation}
R^2\oint_{\partial\Omega}
d\Omega
=R^2\int_0^{2\pi}d\omega
\int_{0}^{\pi}\sin\iota d\iota.
\end{equation}
The above calculation finally gives the final expression of energy flux carried by scalar radiation, 
\begin{equation}
\label{scalar-energy-flux}
\mathcal{F}_{S}
=\frac{32}{5}
\frac{\nu^2x^5}
{(1-e_r^2)^{7/2}}\cdot
\frac{25 }{24576}\Delta^2
\frac{x^2}{(1-e_r^2)^2}
\left[1+\frac{19}{2}e_r^2
+\frac{69}{8}e_r^4
+\frac{9}{16}e_r^6\right].
\end{equation}

Now, we calculate the flux of tensor radiation, which is defined as follows,
\begin{equation}
\label{tensor-energy-flux-definition}
\begin{aligned}
\mathcal{F}_{T}
=\frac{1}{32\pi}R^2
\oint_{\partial\Omega}
\langle\dot{h}^{\rm TT}_{jk}\dot{h}^{\rm TT}_{jk}\rangle d\Omega
=\frac{1}{16\pi}R^2
\oint_{\partial\Omega}
\langle\dot{h}_{+}^2+\dot{h}_{\times}^2\rangle d\Omega
=\frac{1}{4\pi}(\nu m)^2
\oint_{\partial\Omega}
\langle\dot{\xi}_{+}^2
+\dot{\xi}_{\times}^2\rangle d\Omega.
\end{aligned}
\end{equation}
Through some similar mathematical processes, we get
\begin{equation}
\label{tensor-energy-flux}
\mathcal{F}_{T}
=\frac{32}{5}
\frac{\nu^2x^5}
{(1-e_r^2)^{7/2}}
\left[\left(1
+\frac{73}{24}e_r^2
+\frac{37}{96}e_r^4\right)
+\delta\varpi\cdot
\frac{x^2}{(1-e_r^2)^2}
\left(\frac{4}{3}
+\frac{449}{36}e_r^2
+\frac{1195}{144}e_r^4
+\frac{11}{48}e_r^6\right)\right].
\end{equation}
Finally, the total dissipative energy is the sum of scalar flux (\ref{scalar-energy-flux}) and tensor flux (\ref{tensor-energy-flux}), shown as
\begin{equation}
\label{total-energy-flux}
\begin{aligned}
\mathcal{F}\equiv
\mathcal{F}_{S}
+\mathcal{F}_{T}
&=\frac{32}{5}
\frac{\nu^2x^5}
{(1-e_r^2)^{7/2}}
\Bigg\{\left(1
+\frac{73}{24}e_r^2
+\frac{37}{96}e_r^4\right)\\
&\qquad+\frac{x^2}{(1-e_r^2)^2}
\Bigg[\left(\frac{25}{24576}\Delta^2
+\frac{4}{3}\delta\varpi\right)
+\left(\frac{475}{49152}\Delta^2
+\frac{449}{36}\delta\varpi\right)e_r^2\\
&\qquad\qquad+\left(\frac{575}{65536}\Delta^2
+\frac{1195}{144}\delta\varpi\right)e_r^4
+\left(\frac{75}{131072}\Delta^2
+\frac{11}{48}\delta \varpi\right)e_r^6
\Bigg]\Bigg\}.
\end{aligned}
\end{equation}
Setting the coupling to $0$, we get the flux in GR, which can be found in many famous references and textbooks, such as \cite{PetersMathews1963,Will2014,MTW,Maggiore2008,Blanchet2014}. Comparing the leading order, the DCS modification appears in the 2PN order. Taking the eccentricity $e_r$ as $0$, this result is consistent with that obtained by \cite{ZhaoLi2023}. (We find there are some typos in our previous work \cite{ZhaoLi2023}. The coefficient before coupling $\delta\varpi$ in Eq.\,(135) should be $4/3$ rather than $8/3$. This typo influences all the subsequent related coefficients in Eqs.\,(140, 148, 157).)

\subsection{Angular-Momentum Flux}
Because the eccentric orbits are described by two independent parameters (the semimajor axis and eccentricity), the secular evolution is influenced by not only the balance of conserved energy but also the OAM. The angular-momentum flux is also carried by both scalar and tensor radiations. The scalar sector is defined as
\begin{equation}
\mathcal{L}^{k}_S
=-\beta_0R^2\int\epsilon_{ijk}
\langle\dot{\vartheta}
x_i\partial_{j}\vartheta\rangle d\Omega.
\end{equation}
The vector $\mathbf{x}$ is just the direction of the observer,  $\mathbf{x}\equiv\hat{\mathbf{N}}$, and the gradient operator is $\partial_{j}\equiv\partial/\partial x_j$. For simplicity, we define the flux-density vector as $\tau_{S}^k=-\epsilon_{ijk}\dot{\vartheta}x_i\partial_{j}\vartheta$, with components
\begin{equation}
\begin{aligned}
\tau_{S}^x&=-\cot\iota\sin\omega(\partial_{\omega}\vartheta)\dot{\vartheta}+\cos\omega(\partial_\iota\vartheta)\dot{\vartheta},\\
\tau_{S}^y&=-\sin\omega(\partial_\iota\vartheta)\dot{\vartheta}-\cot\iota\cos\omega(\partial_{\omega}\vartheta)\dot{\vartheta},\\
\text{and}\quad\tau_{S}^z&=(\partial_{\omega}\vartheta)\dot{\vartheta},
\end{aligned}
\end{equation}
respectively. Putting the scalar radiation (\ref{vartheta-result}) into this definition and averaging it, we get
\begin{equation}
\label{scalar-flux-density-average}
\begin{aligned}
\beta_0\langle\tau_{S}^x\rangle
&=-\frac{25}{65536}
\frac{(2\nu m)^2}{\pi}
\frac{\xi^{11/2}}{m}\Delta^2
\cdot\frac{8+24e_r^2+3e_r^4}
{(1-e_r^2)^4}\cdot
\sin\iota\cos^3\iota\sin\omega,\\
\beta_0\langle\tau_{S}^y\rangle
&=-\frac{25}{65536}
\frac{(2\nu m)^2}{\pi}
\frac{\xi^{11/2}}{m}\Delta^2
\cdot\frac{8+24e_r^2+3e_r^4}
{(1-e_r^2)^4}\cdot
\sin\iota\cos^3\iota\cos\omega,\\
\beta_0\langle\tau_{S}^z\rangle
&=\frac{25}{65536}
\frac{(2\nu m)^2}{\pi}
\frac{\xi^{11/2}}{m}\Delta^2
\cdot\frac{8+24e_r^2+3e_r^4}
{(1-e_r^2)^4}\cdot
\sin^2\iota\cos^2\iota.
\end{aligned}
\end{equation}
The $x$ and $y$-components are proportional to $\sin\omega$ and $\cos\omega$, respectively, such that integrating them over the full solid angle gives zero. The only non-zero component is $\tau_{S}^{z}$. This is the consequence of the non-precessing assumption. The full solid-angle integral gives the $z$-component of angular-momentum flux carried by scalar radiation, 
\begin{equation}
\label{scalar-angular-momentum-flux}
\mathcal{L}_{S}^{z}
=\frac{32}{5}\frac{m \nu ^2x^{7/2}}
{(1-e_r^2)^2}\cdot
\frac{25}{24576}\Delta^2
\frac{x^2}{(1-e_r^2)^2}
\left(1+3e_r^2+\frac{3}{8}e_r^4\right).
\end{equation}

Now, we turn to the angular-momentum flux by tensor radiation. The definition of flux is 
\begin{equation}
\begin{aligned}
\mathcal{L}_T^k
&=\frac{1}{32\pi}R^2
\int\epsilon^{ijk}\langle
2h_{il}^{\mathrm{TT}} \dot{h}_{jl}^{\mathrm{TT}}
-\dot{h}_{lm}^{\mathrm{TT}}
x_i\partial_jh_{lm}^{\mathrm{TT}}\rangle d\Omega\\
&=\frac{1}{8\pi}(\nu m)^2
\int\epsilon^{ijk}\langle
2\xi_{il}^{\mathrm{TT}} \dot{\xi}_{jl}^{\mathrm{TT}}
-\dot{\xi}_{lm}^{\mathrm{TT}}
x_i\partial_j\xi_{lm}^{\mathrm{TT}}\rangle d\Omega,
\end{aligned}
\end{equation}
and of the flux density is
\begin{equation}
\tau_{T}^k=\epsilon^{ijk}
(2\nu m)^2
(2\xi_{il}^{\mathrm{TT}} \dot{\xi}_{jl}^{\mathrm{TT}}
-\dot{\xi}_{lm}^{\mathrm{TT}}
x_i\partial_j\xi_{lm}^{\mathrm{TT}}),
\end{equation}
with components
\begin{equation}
\begin{aligned}
\tau^x_T&=(2\nu m)^2
\left\{2\cos\omega
\Big[(\partial_{\iota}\xi_{+})
\dot{\xi}_{+}
+(\partial_{\iota}\xi_{\times})
\dot{\xi}_{\times}\Big]
-2\cot\iota\sin\omega
\Big[(\partial_{\omega}\xi_{+})
\dot{\xi}_{+}
+(\partial_{\omega}\xi_{\times})
\dot{\xi}_{\times}\Big]\right\},\\
\tau^y_T&=(2\nu m)^2
\left\{-2\sin\omega
\Big[(\partial_{\iota}\xi_{+})\dot{\xi}_{+}
+(\partial_{\iota}\xi_{\times})
\dot{\xi}_{\times}\Big]
-2\cot\iota\cos\omega
\Big[(\partial_{\omega}\xi_{+})\dot{\xi}_{+}
+(\partial_{\omega}\xi_{\times})
\dot{\xi}_{\times}\Big]\right\},\\
\tau^z_T&=(2\nu m)^2
\left\{12\Big[\dot{\xi}_{\times}\xi_{+}
-\dot{\xi}_{+}\xi_{\times}\Big]
+2\Big[(\partial_{\omega}\xi_{\times})
\dot{\xi}_{\times}
+(\partial_{\omega}\xi_{+})
\dot{\xi}_{+}\Big]\right\}.
\end{aligned}
\end{equation}
We find again that the $x$ and $y$-components contain factors $\cos\omega$ and $\sin\omega$, which leads to zero results after full-solid angle integration. The remained component is 
\begin{equation}
\label{tensor-angular-momentum-flux}
\mathcal{L}_{T}^{z}
=\frac{32}{5}\frac{m \nu ^2x ^{7/2}}{(1-e_r^2)^2}\cdot
\left[\left(1+\frac{7}{8}e_r^2\right)
+\delta\varpi\cdot\frac{x^2}{(1-e_r^2)^2}
\left(\frac{4}{3}+\frac{16}{3}e_r^2
+\frac{35}{48}e_r^4\right)\right].
\end{equation}
Combining Eqs.\,(\ref{scalar-angular-momentum-flux}) and (\ref{tensor-angular-momentum-flux}), we get the total dissipative OAM, which is
\begin{equation}
\label{total-angular-momentum-flux}
\begin{aligned}
\mathcal{L}\equiv\mathcal{L}_{S}^{z}+\mathcal{L}_{T}^{z}
&=\frac{32}{5}\frac{m\nu^2x^{7/2}}{(1-e_r^2)^2}\cdot
\Bigg\{\left(1+\frac{7}{8}e_r^2\right)\\
&+\frac{x^2}{(1-e_r^2)^2}
\left[\left(\frac{25}{24576}\Delta^2
+\frac{4}{3}\delta\varpi\right)
+\left(\frac{25}{8192}\Delta^2
+\frac{16}{3}\delta\varpi\right)e_r^2
+\left(\frac{25}{65536}\Delta^2
+\frac{35}{48}\delta\varpi\right)e_r^4\right]\Bigg\}.
\end{aligned}
\end{equation}
The first term is just the result given by GR at the leading order \cite{Maggiore2008,Will2014}. When setting $e_r$ to zero, the OAM flux is related to energy flux by $\mathcal{F}=\Omega\cdot\mathcal{L}$ \cite{Maggiore2008}.

\subsection{Orbital Evolution}
Without the radiation loss of energy and OAM, the orbital elements, $a_r$ and $e_r$, are two constants related to conserved quantities. However, the energy and OAM dissipation leads to the secular evolution of these elements. This evolution is determined by the balance equations,
\begin{equation}
\label{balance-equation}
\frac{d(\mu\varepsilon)}{d\tau}=-\mathcal{F},\quad\text{and}\quad\frac{d(\mu h)}{dt}=-\mathcal{L},
\end{equation}
where $\mu\equiv m\nu$ is the reduced mass of the BBH system. The left-hand sides of Eq.\,(\ref{balance-equation}) are
\begin{equation}
\frac{d(\mu\varepsilon)}{dt}
=\frac{4}{3}\delta\varpi
\frac{e_rx^3}
{(1-e_r^2)^3}e_{r}'(t)
-\frac{1}{2}\left[1
-2\cdot\delta\varpi\cdot
\frac{x^2}{(1-e_r^2)^2}\right]
x'(t),
\end{equation}
and
\begin{equation}
\frac{dh}{dt}
=-\frac{m}{\sqrt{x}}
\frac{e_r}{\sqrt{1-e_r^2}}
\left[1-\frac{\delta\varpi}{3}
\frac{x^2}{(1-e_r^2)^2}(8+e_r^2)\right]e_{r}'(t)
-\frac{m}{2}
\frac{\sqrt{1-e_r^2}}{x^{3/2}}
\left[1-\frac{\delta\varpi}{2}
\frac{x^2}{(1-e_r^2)^2}(2+e_r^2)\right]x'(t).
\end{equation}
The balance equation (\ref{balance-equation}) gives the independent evolution equation of element $e_r$ and gauge-invariant quantity $x$ (as an equivalent substitutes of the semimajor axis $a_r$),
\begin{equation}
\label{dx-dt}
\begin{aligned}
m\frac{dx}{dt}
&=\frac{64}{5}
\frac{\nu x^5}
{(1-e_r^2)^{7/2}}
\Bigg\{\left(
1+\frac{73}{24}e_r^2
+\frac{37}{96}e_r^4\right)\\
&\quad+\frac{x^2}
{(1-e_r^2)^2}
\Bigg[\left(\frac{25}{24576}\Delta^2
+\frac{10}{3}\delta\varpi\right)
+\left(\frac{475}{49152}\Delta^2
+\frac{43}{3}\delta\varpi\right)e_r^2\\
&\qquad+\left(\frac{575}{65536}\Delta^2
+\frac{133}{18}\delta\varpi\right)e_r^4
+\left(\frac{75}{131072}\Delta^2
+\frac{11}{48}\delta\varpi\right)e_r^6\Bigg]\Bigg\},
\end{aligned}
\end{equation}
and
\begin{equation}
\label{der-dt}
\begin{aligned}
m\frac{de_r}{dt}
&=-\frac{304}{15}
\frac{\nu x^4}
{(1-e_r^2)^{5/2}}\cdot e_r
\Bigg\{\left(1
+\frac{121}{304}e_r^2\right)\\
&\quad+\frac{x^2}{(1-e_r^2)^2}
\Bigg[\left(\frac{375}{155648}\Delta^2+\frac{421}{114}\delta\varpi\right)
+\left(\frac{1125}{311296}\Delta^2
+\frac{907}{228}\delta\varpi\right)e_r^2
+\left(\frac{375}{1245184}\Delta^2
+\frac{143}{456}\delta\varpi\right)e_r^4
\Bigg]\Bigg\}.
\end{aligned}
\end{equation}
It is too hard to directly write down the analytic solution to Eqs.\,(\ref{dx-dt}) and (\ref{der-dt}). An alternative method is to construct the evolution of the eccentricity with the frequency. Dividing Eq.\,(\ref{dx-dt}) by (\ref{der-dt}) gives
\begin{equation}
\frac{dx}{de_r}
=-\frac{12}{19}
\frac{x}{e_r}
\frac{1+\frac{73}{24}e_r^2
+\frac{37}{96}e_r^4}
{(1-e_r^2)\left(1+\frac{121}{304}e_r^2\right)}
\left\{1-\frac{x^2}{1-e_r^2}
\frac{\mathcal{W}_0
+\mathcal{W}_2e_r^2
+\mathcal{W}_4e_r^4
+\mathcal{W}_6e_r^6}
{\left(1+\frac{121}{304}e_r^2\right)
\left(1+\frac{73}{24}e_r^2+\frac{37}{96}e_r^4\right)}\right\},
\end{equation}
with some DCS coefficients listed as following
\begin{equation}
\label{dx-der}
\begin{aligned}
\mathcal{W}_0
&=\frac{325}{233472}\Delta^2
+\frac{41}{114}\delta\varpi,\qquad
\mathcal{W}_2
=\frac{16925}{7471104}\Delta^2
-\frac{245}{2736}\delta\varpi,\Big.\\
\mathcal{W}_4
&=\frac{2325}{1245184}\Delta^2
+\frac{7151}{10944}\delta\varpi,\qquad
\mathcal{W}_6
=\frac{2225}{19922944}\Delta^2
-\frac{649}{21888}\delta\varpi.
\end{aligned}
\end{equation}
The zero-order term is fully consistent with that shown in Ref.\,\cite{Maggiore2008}. The overall minus sign means that the eccentricity decreases from an initial value during the radiation reaction, in which the orbital frequency increases. This effect is generally called orbital circularization \cite{Maggiore2008}. Although the sign of DCS modification cannot be determined for unknown bodies' masses and spins, the coefficients $\mathcal{W}_n$ are always small quantifies, weakly changing the decay rate. Then the eccentricity tends to zero as the frequency increases. DCS theory does not modify this conclusion. 

The above equation (\ref{dx-der}) can be solved perturbatively. Writing the solution as the summation of the GR part and DCS modification
\begin{equation}
x(e_r)=x_0(e_{r})
+\zeta\cdot x_{1}(e_r),
\end{equation}
and putting it into Eq.\,(\ref{dx-der}), the zero-order solution is 
\begin{equation}
\label{x0-solution}
x_0(e_r)=c_0(1-e_r^2)
e_r^{-12/19}
\left[1+\frac{121}{304}e_r^2\right]^{-870/2299},
\end{equation}
while the first-order one is 
\begin{equation}
\label{x1-solution}
\begin{aligned}
x_1(e_r)&=\frac{c_{1}(1-e_r^2)}{e_r^{12/19}(304+121e_r^2)^{870/2299}}+2^{-63/2299}19^{-1740/2299}\frac{c_0^3(1-e_r^2)e_r^{-36/19}}{(121e_r^2+304)^{870/2299}}\\
&\times\Bigg[\left(-\frac{325}{3735552}\Delta^2-\frac{41}{1824}\delta\varpi\right)\text{HyperGeometricF}\left(-\frac{12}{19},\frac{6338}{2299};\frac{7}{19};-\frac{121}{304}e_r^2\right)\\
&+\left(\frac{16925}{69730304}\Delta^2-\frac{35}{3648}\delta\varpi\right)e_r^2\cdot\text{HyperGeometricF}\left(\frac{7}{19},\frac{6338}{2299};\frac{26}{19};-\frac{121}{304}e_r^2\right)\\
&+\left(\frac{6975}{129499136}\Delta^2+\frac{7151}{379392}\delta\varpi\right)e_r^4\cdot\text{HyperGeometricF}\left(\frac{26}{19},\frac{6338}{2299},\frac{45}{19},-\frac{121}{304}e_r^2\right)\\
&+\left(\frac{445}{239075328}\Delta^2-\frac{649}{1313280}\delta\varpi\right)e_r^6\cdot\text{HyperGeometricF}\left(\frac{45}{19},\frac{6338}{2299};\frac{64}{19};-\frac{121}{304}e_r^2\right)\Bigg].
\end{aligned}
\end{equation}
Here, ``HyperGeometricF" is the Hypergeometric function of the first kind \cite{Silverman1972}, and $c_0$, $c_1$ in Eqs.\,(\ref{x0-solution}, \ref{x1-solution}) are two integration constants, determined by a specific initial condition $x(e_r=e_0)=x_{(0)}$, i.e., $x_0(e_0)=x_{(0)}$ and $x_1(e_0)=0$.

\section{\label{sec:Fourier-waveform}Post-Circular Frequency-Domain Waveforms}
This section focuses on the frequency-domain waveforms, requiring expressing $e_r$ as a function of $x$ and writing the time-domain waveforms in terms of another angular variable, mean anomaly $\ell$. However, obtaining the inverse function from Eq.\,(\ref{x1-solution}) is almost impossible. So another necessary approximation, small-eccentricity approximation, is adopted in this procedure. We expand all involved functions in $e_r$ and $e_0$ up to the forth order, $\sim\mathcal{O}(e_0^4)$. This expansion is valid for initial eccentricity less than $0.3$, i.e., $e_0\lesssim0.3$ \cite{Yunes2009ecc}. This scheme will give an analytic calculation on the Fourier waveforms.

\subsection{\label{subsec:eccentricity-Fourier}Frequency-Domain Evolution of Eccentricity}
Firstly, the solution (\ref{x0-solution}, \ref{x1-solution}) can be expanded as
\begin{equation}
\label{x0-er}
x_0(e_r)\simeq x_{(0)}
\left(\frac{e_0}{e_r}\right)^{12/19}
\left[1+\frac{3323}{2888}
(e_0^2-e_r^2)
+\frac{37765681}{33362176}
e_0^4
-\frac{11042329}{8340544}
e_0^2e_r^2
+\frac{6403635}{33362176}e_r^4
\right],
\end{equation}
and
\begin{equation}
\label{x1-er}
\begin{aligned}
x_1(e_r)&\simeq x_{(0)}^3
\left(\frac{e_0}{e_r}\right)^{12/19}
\Bigg\{\Bigg[\tilde{\mathcal{P}}^{(0)}_{-24/19}\left(\frac{e_r}{e_0}\right)^{-24/19}
+\tilde{\mathcal{P}}^{(0)}_{0}\Bigg]\\
&+e_r^2\Bigg[\tilde{\mathcal{P}}^{(2)}_{-24/19}\left(\frac{e_r}{e_0}\right)^{-24/19}
+\tilde{\mathcal{P}}^{(2)}_{0}
+\tilde{\mathcal{P}}^{(2)}_{14/19}\left(\frac{e_r}{e_0}\right)^{14/19}
+\tilde{\mathcal{P}}^{(2)}_{2}\left(\frac{e_r}{e_0}\right)^{2}\Bigg]\\
&+e_r^4\Bigg[
\tilde{\mathcal{P}}^{(4)}_{-24/19}
\left(\frac{e_r}{e_0}\right)^{-24/19}
+\tilde{\mathcal{P}}^{(4)}_{0}
+\tilde{\mathcal{P}}^{(4)}_{14/19}
\left(\frac{e_r}{e_0}\right)^{14/19}
+\tilde{\mathcal{P}}^{(4)}_{2}
\left(\frac{e_r}{e_0}\right)^{2}
+\tilde{\mathcal{P}}^{(4)}_{52/19}
\left(\frac{e_r}{e_0}\right)^{52/19}
+\tilde{\mathcal{P}}^{(4)}_{4}
\left(\frac{e_r}{e_0}\right)^{4}\Bigg]\Bigg\}.
\end{aligned}
\end{equation}
It is easy to check that $x=x_{(0)}$ when taking $e_r$ to $e_0$. For subsequent derivation, we define the frequency normalized by its initial value, 
\begin{equation}
\chi\equiv\frac{\Omega}{\Omega_0}=\left[\frac{x}{x_{(0)}}\right]^{3/2}\equiv\chi(e_r).
\end{equation}
This quantity can also be regarded as $F/F_0$, with $F, F_0$ being $F\equiv\Omega/2\pi$ and its initial value as used in Ref.\,\cite{Yunes2009ecc}. 

The zero-order expression in terms of $\chi$ from Eq.\,(\ref{x0-er}) is
\begin{equation}
\label{chi0-solution}
\chi_0\simeq
\left(\frac{e_0}{e_r}\right)^{18/19}
\left[1+\frac{9969}{5776}
(e_0^2-e_r^2)
+\frac{73212015}{33362176}
e_0^4
-\frac{99380961}{33362176}
e_0^2e_r^2
+\frac{13084473}{16681088}
e_r^4\right],
\end{equation}
and the first-order expression from Eq.\,(\ref{x1-er}) is
\begin{equation}
\label{chi1-solution}
\begin{aligned}
\chi_1&\simeq x_{(0)}^2
\left(\frac{e_0}{e_r}\right)^{18/19}
\Bigg\{\Bigg[\mathcal{P}^{(0)}_{0}
+\mathcal{P}^{(0)}_{-24/19}\left(\frac{e_r}{e_0}\right)^{-24/19}\Bigg]\\
&+e_r^2\Bigg[\mathcal{P}^{(2)}_{0}
+\mathcal{P}^{(2)}_{-24/19}
\left(\frac{e_r}{e_0}\right)^{-24/19}
+\mathcal{P}^{(2)}_{-2}
\left(\frac{e_r}{e_0}\right)^{-2}
+\mathcal{P}^{(2)}_{-62/19}
\left(\frac{e_r}{e_0}\right)^{-62/19}
\Bigg]\\
&+e_r^4\Bigg[\mathcal{P}^{(4)}_{0}
+\mathcal{P}^{(4)}_{-24/19}
\left(\frac{e_r}{e_0}\right)^{-24/19}
+\mathcal{P}^{(4)}_{-2}
\left(\frac{e_r}{e_0}\right)^{-2}
+\mathcal{P}^{(4)}_{-62/19}
\left(\frac{e_r}{e_0}\right)^{-62/19}\\
&\qquad\qquad\qquad+\mathcal{P}^{(4)}_{-4}
\left(\frac{e_r}{e_0}\right)^{-4}
+\mathcal{P}^{(4)}_{-100/19}
\left(\frac{e_r}{e_0}\right)^{-100/19}
\Bigg]\Bigg\}.
\end{aligned}
\end{equation}
The coefficients involved in Eq.\,(\ref{chi1-solution}) are listed in Appendix \ref{Appendix-A}. One can inversely solve Eqs.\,(\ref{chi0-solution}, \ref{chi1-solution}) to express the eccentricity as a function of normalized frequency,
\begin{equation}
\label{e-chi}
\begin{aligned}
e_r=e_r(\chi)
&=e_0\chi^{-19/18}\Bigg[1
+\frac{3323}{1824}e_0^2\left(1-\chi^{-19/9}\right)\\
&~~~~~~ +\frac{15994231}{6653952}e_0^4\left(1-\frac{66253974}{15994231}\chi^{-19/9}+\frac{50259743}{15994231}\chi^{-38/9}\right)
+\zeta\left(\delta\mathcal{E}_{0}
+\delta\mathcal{E}_{2}e_0^2
+\delta\mathcal{E}_{4}e_0^4
\right)\Bigg].
\end{aligned}
\end{equation}
The leading-order terms have been shown in Eq.\,(\ref{e-chi}) and the DCS modifications are
\begin{equation}
\label{e-chi-coefficient}
\begin{aligned}
\delta\mathcal{E}_{0}&=\chi^{-19/18}\left[\mathcal{S}^{(0)}_{0}+\mathcal{S}^{(0)}_{4/3}\chi^{4/3}\right],\\
\delta\mathcal{E}_{2}&=\chi^{-19/18}\left[\mathcal{S}^{(2)}_{-19/9}\chi^{-19/9}+\mathcal{S}^{(2)}_{-7/9}\chi^{-7/9}+\mathcal{S}^{(2)}_{0}+\mathcal{S}^{(2)}_{4/3}\chi^{4/3}\right],\\
\delta\mathcal{E}_{4}&=\chi^{-19/18}\left[\mathcal{S}^{(4)}_{-38/9}\chi^{-38/9}+\mathcal{S}^{(4)}_{-26/9}\chi^{-26/9}+\mathcal{S}^{(4)}_{-19/9}\chi^{-19/9}+\mathcal{S}^{(4)}_{-7/9}\chi^{-7/9}+\mathcal{S}^{(4)}_{0}+\mathcal{S}^{(4)}_{4/3}\chi^{4/3}\right].
\end{aligned}
\end{equation}
The expansion coefficients in Eq.\,(\ref{e-chi-coefficient}) can be found in Appendix \ref{Appendix-B}. The GR sector in Eq.\,(\ref{e-chi}) is consistent with the results in Ref.\,\cite{Yunes2009ecc} and the DCS sector is obtained by this work for the first time. In this way, we directly express the frequency-domain evolution instead of time evolution. This will play an important role in the next calculation of the waveforms in frequency space. 

\subsection{Waveform in Terms of Mean Anomaly}
The so-called mean anomaly is defined by the modified Keplerian equation $\ell(u)\equiv\phi/K=u-e_t\sin u$ (\ref{azimuth-results}). It grows monotonically over time. Therefore, without considering the radiation reaction, it is feasible to express the mean anomaly through orbital frequency $F$, as $\ell=Ft(u)$, bringing great convenience in calculating Fourier waveforms. To complete this calculation, we firstly re-express Eqs.\,(\ref{Newtonian-polarization-in-rphi}, \ref{DCS-polarization-in-rphi}) in terms of azimuth and eccentric anomaly. The Newtonian approximation is
\begin{equation}
\label{Newtonian-polarization-u-phi}
\begin{aligned}
\xi_{+}^{(0)}
&=\frac{\xi}{2}\frac{e_r\cos u}
{1-e_r\cos u}\sin^2\iota
-\frac{\xi}{2}\frac{1}{(1-e_r\cos u)^2}(1+\cos^2\iota)\\
&\qquad\times
\Big\{[2(1-e_r^2)-e_r\cos u
+e_r^2\cos^2u]
\cos(2\phi+2\omega)
+2e_r\sqrt{1-e_r^2}\sin u
\sin(2\phi+2\omega)\Big\},\\
\xi_{\times}^{(0)}
&=\frac{\xi\cos\iota}
{(1-e_r\cos u)^2}
\Big\{2e_r\sqrt{1-e_r^2}
\sin u\cos(2\phi+2\omega)
-[2(1-e_r^2)-e_r\cos u
+e_r^2\cos^2u]
\sin(2\phi+2\omega)\Big\}.
\end{aligned}
\end{equation}
The DCS modification is
\begin{equation}
\label{DCS-polarization-u-phi}
\begin{aligned}
\delta\xi_{+}
&=\frac{\delta\varpi}{6}
\frac{\xi^3}{(1-e_r^2)^{3/2}}
\frac{1}{(1-e_r\cos u)^3}
\sin^2\iota\Big[
e_r^2-3(e_r\cos u)
+3(e_r\cos u)^2
-(e_r\cos u)^3\Big]\\
&\qquad+\frac{\delta\varpi}{6}
\frac{\xi^3}{(1-e_r^2)^{3/2}}
\frac{1}{(1-e_r\cos u)^3}
(1+\cos^2\iota)
\Big\{2\Big[(1+e_r^2)
(e_r\cos u)-2e_r^2\Big](e_r\sin u)
\sin(2\phi+2\omega)\\
&\qquad\qquad-\sqrt{1-e_r^2}
\Big[6+e_r^2
-(3+2e_r^2)(e_r\cos u)
-3(e_r\cos u)^2
+(e_r\cos u)^3\Big]
\cos(2\phi+2\omega)\Big\},\\
\delta\xi_{\times}
&=\frac{\delta \varpi}{3}
\frac{\xi^3}{(1-e_r^2)^{3/2}}
\frac{1}{(1-e_r\cos u)^3}
\cos\iota
\Big\{2\Big[2e_r^2
-(1+e_r^2)(e_r\cos u)\Big]
(e_r\sin u)
\cos(2\phi+2\omega)\\
&\qquad-\sqrt{1-e_r^2}
\Big[6+e_r^2
-(3+2e_r^2)(e_r\cos u)
-3(e_r\cos u)^2
+(e_r\cos u)^3\Big]
\sin(2\phi+2\omega)\Big\}.
\end{aligned}
\end{equation}
The transformation from eccentric anomaly to mean anomaly is the famous solution to the modified Keplerian equation (\ref{azimuth-results}) by an infinite Bessel expansion \cite{Boetzel2017,Roy2020}, which is 
\begin{equation}
\label{u-ell-relation}
\begin{aligned}
u-\ell&=\sum_{s=0}^{\infty}\frac{2}{s}J_{s}(se_t)\sin(s\ell)\\
&\simeq\left[1+\frac{1}{8}e_r^2
+\delta\varpi\cdot\xi^2\left(
-\frac{1}{3}
-\frac{11}{24}e_r^2
\right)\right]e_r\sin(\ell)\\
&\qquad+\frac{1}{2}\left[1
+\frac{1}{3}e_r^2
+\delta\varpi\cdot\xi^2
\left(-\frac{2}{3}
-\frac{10}{9}e_r^2\right)\right]
e_r^2\sin(2\ell)\\
&\qquad+\frac{3}{8}\left(1
-\delta\varpi\cdot\xi^2\right)
e_r^3\sin(3\ell)
+\frac{1}{3}\left(1
-\frac{4}{3}\delta\varpi\cdot\xi ^2\right)e_r^4\sin(4\ell).
\end{aligned}
\end{equation}
$J_{s}(z)$ is the $s$-order Bessel function of the first kind. The asymptotic behaviour of the Bessel function for fixed $s$ and small $e_t$ is $\propto e_{t}^{s}$, such that the higher-order Bessel functions are dropped in our small-eccentricity scheme. Another relevant angular variable, azimuth, in Eqs.\,(\ref{Newtonian-polarization-u-phi}, \ref{DCS-polarization-u-phi}), is also regarded as $\phi=(K/2\pi)v=(1+\beta)v$, where $v$ is defined in Eq.\,(\ref{azimuth-results}). Up to the order of $\sim\mathcal{O}(\zeta)$ and $\sim\mathcal{O}(e_r^4)$, using above definition, we have
\begin{equation}
\label{v-u-relation}
\begin{aligned}
v&=u+\left[1+\frac{1}{4}e_r^2
+\delta\varpi\cdot\xi^2
\left(\frac{1}{3}
+\frac{7}{12}e_r^2
\right)\right]e_r\sin(u)
+\frac{1}{4}\left[1
+\frac{1}{2}e_r^2
+\delta\varpi\cdot\xi^2
\left(\frac{2}{3}
+\frac{4}{3}e_r^2
\right)\right]e_r^2\sin(2u)\\
&\qquad+\frac{1}{12}\left(
1+\delta\varpi\cdot\xi^2\right)
e_r^3\sin(3u)
+\frac{1}{32}\left(
1+\frac{4}{3}\delta\varpi
\cdot\xi^2\right)e_r^4\sin(4u),
\end{aligned}
\end{equation}
where the ``angular" eccentricity are written as Eq.\,(\ref{et-ephi-er}). Substituting Eq.\,(\ref{u-ell-relation}) into (\ref{v-u-relation}), we get
\begin{equation}
\label{v-ell-relation}
\begin{aligned}
v&=\ell+\left[1
-\frac{1}{8}e_r^2
+\frac{1}{6}
\delta\varpi\cdot\xi^2e_r^2
\right]e_r\sin(\ell)
+\frac{5}{4}\left[1
+\frac{11}{30}e_r^2
+\delta\varpi\cdot\xi^2
\left(-\frac{2}{15}
+\frac{4}{15}e_r^2
\right)\right]e_r^2\sin(2\ell)\\
&\qquad+\frac{13}{12}\left[
1-\frac{4}{13}\delta\varpi\cdot\xi^2\right]
e_r^3\sin(3\ell)
+\frac{103}{96}\left[
1-\frac{52}{103}\delta\varpi
\cdot\xi^2\right]e_r^4\sin(4\ell),
\end{aligned}
\end{equation}
Finally, combining Eq.\,(\ref{u-ell-relation}) with (\ref{v-ell-relation}), the waveforms in Eqs.\,(\ref{Newtonian-polarization-u-phi}, \ref{DCS-polarization-u-phi}) are rewritten as
\begin{equation}
\begin{aligned}
\xi_{+}^{(0)}
&=x\left\{\frac{1}{48}e_r\Big[
3(8-e_r^2)\sin^2\iota
+4(1+\cos^2\iota)(9-4e_r^2)
\cos(2\beta\ell+2\omega)\Big]
\cos(\ell)\right.\\
&+\frac{1}{24}\Big[
e_r^2(12-4e_r^2)\sin^2\iota
-3(1+\cos^2\iota)
(8-20e_r^2+11e_r^4)
\cos(2\beta\ell+2\omega)\Big]
\cos(2\ell)\\
&+\frac{9}{32}e_r
\Big[2e_r^2\sin^2\iota
-(1+\cos^2\iota)(8-19e_r^2)
\cos(2\beta\ell+2\omega)
\Big]\cos(3\ell)\\
&+\frac{2}{3}e_r^2 \Big[e_r^2\sin^2\iota
-3(1+\cos^2\iota)(2-5e_r^2)
\cos(2\beta\ell+2\omega)\Big]
\cos(4\ell)\\
&-\frac{625}{96}
(1+\cos^2\iota)e_r^3 \cos(2\beta\ell+2\omega)
\cos(5\ell)
-\frac{81}{8}
(1+\cos^2\iota)e_r^4 \cos(2\beta\ell+2\omega)
\cos(6\ell)\\
&-\frac{1}{48}(1+\cos^2\iota) e_r(36-23e_r^2) \sin(2\beta\ell+2\omega)
\sin(\ell)
+\frac{1}{2} (1+\cos^2\iota)
(2-5e_r^2+3e_r^4)
\sin(2\beta\ell+2\omega)
\sin(2\ell)\\
&+\frac{9}{32} 
(1+\cos^2\iota)e_r
(8-19e_r^2)
\sin(2\beta\ell+2\omega)
\sin(3\ell)
+2(1+\cos^2\iota)e_r^2
(2-5e_r^2)
\sin(2\beta\ell+2\omega)
\sin(4\ell)\\
&\left.+\frac{625}{96} (1+\cos^2\iota)e_r^3
\sin(2\beta\ell+2\omega)
\sin(5\ell)
+\frac{81}{8} (1+\cos^2\iota)e_r^4
\sin(2\beta\ell+2\omega)
\sin(6\ell)\right\},
\end{aligned}
\end{equation}
\begin{equation}
\begin{aligned}
\xi_{\times}^{(0)}
&=x\left\{\frac{1}{6}
e_r(9-4e_r^2)\cos\iota
\sin(2\beta\ell+2\omega)
\cos(\ell)
-\frac{1}{4}
(8-20e_r^2+11e_r^4)\cos\iota
\sin(2\beta\ell+2\omega)
\cos(2\ell)\right.\\
&-\frac{9}{16}
e_r(8-19e_r^2)\cos\iota
\sin(2\beta\ell+2\omega)
\cos(3\ell)
-4e_r^2(2-5e_r^2)\cos\iota
\sin(2\beta\ell+2\omega)
\cos(4\ell)\\
&-\frac{625}{48}
e_r^3\cos\iota
\sin(2\beta\ell+2\omega)
\cos(5\ell)
-\frac{81}{4}
e_r^4\cos\iota
\sin(2\beta\ell+2\omega)
\cos(6\ell)\\
&+\frac{1}{24}
e_r(36-23e_r^2)\cos\iota
\cos(2\beta\ell+2\omega)
\sin(\ell)
-(2-5e_r^2+3e_r^4)\cos\iota
\cos(2\beta\ell+2\omega)
\sin(2\ell)\\
&-\frac{9}{16}
e_r(8-19e_r^2)\cos\iota
\cos(2\beta\ell+2\omega)
\sin(3\ell)
-4e_r^2(2-5e_r^2)\cos\iota
\cos(2\beta\ell+2\omega)
\sin(4\ell)\\
&\left.-\frac{625}{48}
e_r^3\cos\iota
\cos(2\beta\ell+2\omega)
\sin(5\ell)
-\frac{81}{4}e_r^4\cos\iota
\cos(2\beta\ell+2\omega)
\sin(6\ell)\right\},
\end{aligned}
\end{equation}
\begin{equation}
\begin{aligned}
\delta\xi_{+}
&=\delta\varpi\cdot x^3\left\{\frac{1}{18}e_r \Big[-(12+15e_r^2)\sin^2\iota
+(1+\cos^2\iota)
(45+41e_r^2)
\cos(2\beta\ell+2\omega)\Big]
\cos(\ell)\right.\\
&-\left[e_r^2\left(1+\frac{224}{288}e_r^2\right)\sin^2\iota
+\frac{1}{96}
(1+\cos^2\iota)
(64-768e_r^2-575e_r^4)
\cos(2\beta\ell+2\omega)\right]
\cos(2\ell)\\
&-\frac{3}{2}e_r\Big[e_r^2 \sin^2\iota
+(1+\cos^2\iota)(3-10e_r^2)
\cos(2\beta\ell+2\omega)\Big]
\cos(3\ell)\\
&-\frac{1}{9}e_r^2\Big[
20e_r\sin^2\iota
+3(1+\cos^2\iota)
(35-83 e_r^2)
\cos(2\beta\ell+2\omega)\Big]
\cos(4\ell)\\
&-\frac{425}{18} (1+\cos^2\iota)e_r^3 \cos(2\beta\ell+2\omega)
\cos(5\ell)
-\frac{1365}{32} (1+\cos^2\iota)e_r^4 \cos(2\beta\ell+2\omega)
\cos(6\ell)\\
&-\frac{1}{18} (1+\cos^2\iota)e_r
(45+31e_r^2) \sin(2\beta\ell+2\omega)
\sin(\ell)
+\frac{1}{96} (1+\cos^2\iota)
(64-768e_r^2-529e_r^4) \sin(2\beta\ell+2\omega)
\sin(2\ell)\\
&+\frac{3}{2}(1+\cos^2\iota)
(3-10e_r^2)e_r \sin(2\beta\ell+2\omega)
\sin(3\ell)
+\frac{1}{3} (1+\cos^2\iota)e_r^2
(35-83e_r^2) \sin(2\beta\ell+2\omega)
\sin(4\ell)\\
&\left.+\frac{425}{18} (1+\cos^2\iota)e_r^3 \sin(2\beta\ell+2\omega)\sin(5\ell)
+\frac{1365}{32} (1+\cos^2\iota)e_r^4 \sin(2\beta\ell+2\omega)
\sin(6\ell)\right\},
\end{aligned}
\end{equation}
and
\begin{equation}
\begin{aligned}
\delta\xi_{\times}
&=\delta\varpi\cdot x^3\left\{\frac{1}{9}e_r (45+41e_r^2)\cos\iota
\sin(2\beta\ell+2\omega)
\cos(\ell)
-\frac{1}{48}
(64-768e_r^2-575e_r^4)
\cos\iota
\sin(2\beta\ell+2\omega)
\cos(2\ell)\right.\\
&-3e_r(3-10e_r^2)\cos\iota \sin(2\beta\ell+2\omega)
\cos(3\ell)
-\frac{2}{3}e_r^2
(35-83e_r^2)\cos\iota \sin(2\beta\ell+2\omega)
\cos(4\ell)\\
&-\frac{425}{9}e_r^3\cos\iota
\sin(2\beta\ell+2\omega)
\cos(5\ell)
-\frac{1365}{16}
e_r^4\cos\iota \sin(2\beta\ell+2\omega)
\cos(6\ell)\\
&+\frac{1}{9}e_r
(45+31e_r^2)\cos\iota \cos(2\beta\ell+2\omega)
\sin(\ell)
-\frac{1}{48}\cos\iota
(64-768e_r^2-529e_r^4) \cos(2\beta\ell+2\omega)
\sin(2\ell)\\
&-3e_r\cos\iota(3-10e_r^2) \cos(2\beta\ell+2\omega)
\sin(3\ell)
-\frac{2}{3}
e_r^2(35-83e_r^2)\cos\iota \cos(2\beta\ell+2\omega)
\sin(4\ell)\\
&\left.-\frac{425}{9}e_r^3\cos\iota \cos(2\beta\ell+2\omega)
\sin(5\ell)
-\frac{1365}{16}e_r^4\cos\iota
\cos(2\beta\ell+2\omega)
\sin(6\ell)\right\}.
\end{aligned}
\end{equation}
So far, we have provided the expressions for the GW polarizations using the mean anomaly.

\subsection{Overview on Frequency-Domain Waveform}
\label{overview-Fourier}
The detected signal is the linear combination of two different polarization modes, 
\begin{equation}
\label{strain-sin-cos-overview}
\begin{aligned}
h(t)=h_{+}F_{+}+h_{\times}F_{\times}
&=\frac{2\nu m}{R}
\sum_{n=0}^{\infty}
\sum_{k=0}^{\infty}
\Bigg\{
\left[A^{(n,k)}_{SS}(\xi,e_r)
\sin(n\ell)
+A^{(n,k)}_{CS}(\xi,e_r)
\cos n\ell\right]\sin(k\beta\ell)\\
&\qquad\qquad\qquad\qquad
+\left[A^{(n,k)}_{SC}(\xi,e_r)
\sin(n\ell)
+A^{(n,k)}_{CC}(\xi,e_r)
\cos n\ell\right]
\cos(k\beta\ell)\Bigg\},
\end{aligned}
\end{equation}
where $h_{+,\times}=(2\nu m/R)\xi_{+,\times}$ and $F_{+,\times}$ are the pattern functions of the GW detectors. $A^{(n,k)}_{SS}, A^{(n,k)}_{CS}, A^{(n,k)}_{SC}, A^{(n,k)}_{CC}$ are some coefficients including semimajor axis $\xi$ and eccentricity $e_r$. $n$ is a non-negative integer and $k$ represents the waveform modulation from the precession rate. And the frequency and eccentricity are both functions of the time, i.e., $F=F(t)$ and $e_r=e_r(t)$. For convenience, we transform Eq.\,(\ref{strain-sin-cos-overview}) to the following form,
\begin{equation}
\label{strain-exp-overview}
h(t)=\frac{2\nu m}{R}
\sum_{nk}A_{nk}(\xi,e_r)e^{-i(n+k\beta)\ell},\quad\text{with}\quad
\sum_{nk}\equiv
\sum_{n=-\infty}^{\infty}
\sum_{k=-\infty}^{\infty},
\end{equation}
and the Fourier transformation of Eq.\,(\ref{strain-exp-overview}) is given by
\begin{equation}
\label{Fourier-overview}
\tilde{h}(f)=\int_{-\infty}^{\infty}
h(t)e^{i2\pi ft}dt
=\frac{2\nu m}{R}
\sum_{nk}\int_{-\infty}^{\infty}
A_{nk}(F)
e^{-i(n+k\beta)\ell}
e^{i2\pi ft}dt,
\end{equation}
The above integration can be calculated by stationary phase approximation (SPA) \cite{Maggiore2008,XingZhang2017BD,TanLiu2020BD,Klein2018,Tessmer2013}. The final result is just contributed by the terms calculating stationary points,
\begin{equation}
\label{SPA-overview}
\begin{aligned}
\tilde{h}(f)
&\simeq\frac{2\nu m}{R}
\tilde{\sum_{nk}}
\int_{-\infty}^{\infty}
A_{nk}(F)e^{i[2\pi ft-(n+k\beta)\ell]}dt\\
&=\frac{2\nu m}{R}
\tilde{\sum_{nk}}
A_{nk}(F_{nk})
\sqrt{\frac{2\pi}{\ddot{\psi}_{nk}}}
\exp\left\{i\left[2\pi ft_{nk}
-(n+k\beta_{nk})\ell_{nk}-\frac{\pi}{4}\right]\right\},
\end{aligned}
\end{equation}
where
\begin{equation}
\label{ddot-psi}
\ddot{\psi}_{nk}
=(n+k\beta_{nk})\ddot{\ell}_{nk}
+2k\dot{\beta}_{nk}\dot{\ell}_{nk}
+k\ddot{\beta}_{nk}\ell_{nk}.
\end{equation}
The new operator $\tilde{\sum}_{nk}$ means summing all the terms involving stationary points, which is determined by 
\begin{equation}
\label{stationary-point-equation-overview}
f=(n+k\beta_{nk})\cdot F_{nk}.
\end{equation}
The periastron advance brings some difficulties in calculating the stationary points because $\beta$ is also the function of eccentricity and semimajor axis, which further the functions of frequency under radiation reaction. One can solve the above equation (\ref{stationary-point-equation-overview}) perturbatively,
\begin{equation}
\label{stationary-point-overview}
F_{nk}=F_{nk}(f)
\approx\frac{f}{n}
\Big[1+\delta F_{nk}(f)\Big].
\end{equation}
The stationary frequencies in the Newtonian case are $f/n$, and $\delta F_{nk}$ is the DCS modification. The mean anomaly $\ell=Ft(u)$ should be rewritten as
\begin{equation}
\label{mean-anomaly-overview}
\ell(F)=2\pi\int_{0}^{t}Fdt
=2\pi\int_{F_0}^{F}
(F/\dot{F})dF
=\ell_{c}+2\pi\int(F/\dot{F})dF,
\end{equation}
involving the radiation reaction. Its first- and second-order derivatives of the mean anomaly are given by
\begin{equation}
\label{derivative-mean-anomaly-overview}
\dot{\ell}=\dot{F}\frac{d\ell}{dF}=\dot{\ell}(F),
\quad\text{and}\quad
\ddot{\ell}=\ddot{F}\frac{d\ell}{dF}+\dot{F}^2\frac{d^2\ell}{dF^2},
\end{equation}
respectively, where
\begin{equation}
\label{derivative-F}
\dot{F}=\frac{dF}{de_r}\frac{de_r}{dt}
,\quad\text{and}\quad
\ddot{F}=\dot{F}\frac{d\dot{F}}{dF}.
\end{equation}
Compared with the calculation in the quasi-circular case, the expressions of $\dot{\ell}$ and $\ddot{\ell}$ are eventually expressed as the function of the orbital frequency rather than time. At the stationary points, we have $F_{nk}=F_{nk}(f)$ and then give $\ell_{nk}=\ell(F_{nk})$, $\dot{\ell}_{nk}=\dot{\ell}(F_{nk})$, and $\ddot{\ell}_{nk}=\ddot{\ell}(F_{nk})$. The derivatives of the precession rate are also obtained similarly,
\begin{equation}
\label{derivative-beta}
\dot{\beta}=\dot{F}
\frac{d\beta}{dF}
=\dot{\beta}(F),
\quad\text{and}\quad
\ddot{\beta}
=\ddot{F}
\frac{d\beta}{dF}
+\dot{F}^2
\frac{d^2\beta}{dF^2}.
\end{equation}
So at the stationary point, the precession rate and its derivatives are calculated as $\beta_{nk}=\beta(F_{nk})$, $\dot{\beta}_{nk}=\dot{\beta}(F_{nk})$, and $\ddot{\beta}_{nk}=\ddot{\beta}(F_{nk})$. In the end, the time is calculated through
\begin{equation}
\label{time}
t=\int_{F_0}^{F}\dot{F}^{-1}dF
=t_{c}+\int\dot{F}^{-1}dF,
\end{equation}
and given by $t_{nk}=t(F_{nk})$ at the stationary point. 

Combining the results $t_{nk}$, $\beta_{nk}$, $\ell_{nk}$, and $\ddot{\psi}_{nk}$ (\ref{ddot-psi}), the frequency-domain waveform takes following form,
\begin{equation}
\label{strain-frequency-overview}
\tilde{h}(f)=\tilde{\sum_{nk}}\mathcal{A}_{nk}(f)e^{i\Psi_{nk}(f)},
\end{equation}
The modified amplitude in Eq.\,(\ref{strain-frequency-overview}) is
\begin{equation}
\label{modified-amplitude-overview}
\mathcal{A}_{nk}=\frac{2\nu m}{R}\cdot\sqrt{2\pi}\cdot
A_{nk}(F_{nk})\cdot\ddot{\psi}_{nk}^{-1/2},
\end{equation}
and the modified phase is
\begin{equation}
\label{modified-phase-overview}
\Psi_{nk}=2\pi ft_{nk}-(n+k\beta_{nk})\ell_{nk}
-\frac{\pi}{4}.
\end{equation}
Here we have completed the overview of the calculation of Fourier waveforms. The corresponding detailed results are reported in the following subsection.

\subsection{Results}
Now, we present the detailed results at the post-circular approximation or in the small-eccentricity limit. The considered waveforms up to order $\sim\mathcal{O}(\zeta)$ and $\sim\mathcal{O}(e_r^4)$ is
\begin{equation}
\label{strain-exp}
h(t)=\frac{2\nu m}{R}\ 
\sum_{n'k'}
A_{nk}(\xi,e_r)
e^{-i(n+k\beta)\ell},
\end{equation}
The re-defined summing operator represents the sum of 26 terms with $n\in  n'\equiv\{-6,-5,\cdots,6\}$ and $k\in k'\equiv\{-2,2\}$. The amplitudes are separated into GR part and DCS modification, i.e., $A_{nk}=\bar{A}_{nk}+\delta A_{nk}$. All these terms are listed in the Appendix \ref{Appendix-C}. The Fourier transformation and SPA approximation give
\begin{equation}
\label{SPA}
\tilde{h}(f)
\simeq\frac{2\nu m}{R}
\tilde{\sum_{n'k'}}
A_{nk}(F_{nk})
\sqrt{\frac{2\pi}{\ddot{\psi}_{nk}}}
\exp\left\{i\left[2\pi ft_{nk}
-(n+k\beta_{nk})\ell_{nk}-\frac{\pi}{4}\right]\right\},
\end{equation}
which sums the 12 terms including the stationary points with $n>0$. $\ddot{\psi}_{nk}$ is defined in Eq.\,(\ref{ddot-psi}). Following the steps provided in Sec \ref{overview-Fourier}, one eventually arrives at the modified amplitude and phase defined in Eqs.\,(\ref{strain-frequency-overview}, \ref{modified-amplitude-overview}, \ref{modified-phase-overview}). For simplicity, the detailed calculation process is shown in appendix \ref{Appendix-D}. The final modified phase is obtained using Eqs.\,(\ref{t-nk}, \ref{beta-nk}, \ref{ell-nk}),
\begin{equation}
\label{Psi-nk}
\begin{aligned}
\Psi_{nk}&=-n\ell_c+2\pi ft_c-\frac{\pi}{4}+\frac{3}{256}n\tilde{u}_f^{-5}\left\{1-\frac{2355}{1462}e_0^2\chi_f^{-19/9}+e_0^4\left(\frac{5222765}{998944}\chi_f^{-38/9}-\frac{2608555}{444448}\chi_f^{-19/9}\right)\right\}\\
&+\frac{3}{256}n\tilde{u}_f^{-5}\cdot\frac{\tilde{u}_f^4}{\nu^{4/5}}\Bigg\{\Bigg[-\frac{125}{12288}+e_0^2\left(\frac{220625}{33521664}\chi_f^{-19/9}-\frac{255125}{71860224}\chi_f^{-31/9}\right)\\
&+e_0^4\left(\frac{733136875}{30571757568}\chi_f^{-19/9}-\frac{8202445375}{458755670016}\chi_f^{-31/9}-\frac{43995625}{1608155136}\chi_f^{-38/9}+\frac{1697398625}{73650143232}\chi_f^{-50/9}\right)\Bigg]\Delta^2\\
&+\Bigg[\frac{8}{3}\frac{k}{n}-\frac{100}{3}-\frac{256}{3}\frac{k}{n}\chi_f^{5/3}(2\pi\mathcal{M}F_0)^{5/3}\ell_c\\
&+e_0^2\left(-\frac{241}{102}\frac{k}{n}\chi_f^{-19/9}+\frac{126745}{16368}\chi_f^{-19/9}-\frac{32185}{35088}\chi_f^{-31/9}-\frac{512}{3}\frac{k}{n}\chi_f^{-4/9}(2\pi\mathcal{M}F_0)^{5/3}\ell_c\right)\\
&+e_0^4\Bigg(\left(\frac{421173635}{14927616}-\frac{800843}{93024}\frac{k}{n}\right)\chi_f^{-19/9}+\left(\frac{22659965}{2465136}\frac{k}{n}-\frac{107411761}{6281856}\right)\chi_f^{-38/9}-\frac{1695694985}{224001792}\chi_f^{-31/9}\\
&\qquad+\frac{214133365}{35961984}\chi_f^{-50/9}+\left(\frac{62560}{171}\chi_f^{-23/9}-\frac{106336}{171}\chi_f^{-4/9}\right)\frac{k}{n}(2\pi\mathcal{M}F_0)^{5/3}\ell_c\Bigg)\Bigg]\delta\varpi\Bigg\},
\end{aligned}
\end{equation}
where
\begin{equation}\label{fdep}
\tilde{u}_f\equiv(2\pi\mathcal{M}f/n)^{1/3},\quad\chi_{f}\equiv(1/n)(f/F_0).
\end{equation}
The chirp mass is defined as
\begin{equation}
\mathcal{M}\equiv m\nu^{3/5}.
\end{equation}
From Eqs.\,(\ref{beta-nk}, \ref{dot-beta-nk}, \ref{ddot-beta-nk}, \ref{ell-nk}, \ref{dot-ell-nk}, \ref{ddot-beta-nk}), the modified amplitudes in Eq.\,(\ref{modified-amplitude-overview}) are obtained, which are listed in Appendix \ref{Appendix-D}. The first line of Eq.\,(\ref{Psi-nk}) has been obtained by \cite{Yunes2009ecc}, but our results differ by a minus sign from theirs, because of the different definitions. Our $\Psi_{nk}$ corresponds to $-i(\pi/4+\Psi_{n})$ in Eq.\,(4.29) of \cite{Yunes2009ecc}. We can recall that the DCS theory modifies the gravitational radiation and reaction at 2PN-order approximation. And because of the existence of periastron advance, the second derivative of phase function in the SPA formula should be replaced by Eq.\,(\ref{ddot-psi}), where $\ell$ is involved. So there are some terms including mean anomaly at the coalescence moment, $\ell_{c}$.

Equation.\,(\ref{Psi-nk}) and Appendix \ref{Appendix-D}
provide the explicit expression of modified waveform \eqref{strain-frequency-overview}. The waveform depends on the following parameters: the total mass $m$, the mass ratio $\nu$, the spins $\bm{S}_{A}$, the merging time $t_c$, the azimuth angle $\omega$, the inclination angle $\iota$ of the observer, the merging phase $\ell_c$, initial frequency $F_{0}$, initial eccentricity $e_0$ ($\lesssim0.3$), and coupling parameters $\zeta$, which is encoded in $\delta\varpi$ and $\Delta^2$, given by Eqs.\,\eqref{DCS-coefficient} and \eqref{delta2}, respectively. The frequency-dependent combinations $\chi_f$ and $\tilde{u}_f$ are given by Eq.\,\eqref{fdep}. The modified phase and amplitudes are two sets of functions of frequency $f$ with two indices $n$ and $k$. The summation in the waveform \eqref{strain-frequency-overview} over $n$ and $k$ is within the ranges $n'\equiv\{-6,-5,\cdots,6\}$ and $k'\equiv\{-2,2\}$.

As stated by Ref.\,\cite{Alexander2018}, for circular-orbit binaries (setting $e_r=0, n=2$, and $k=0$ in Eq.\,(\ref{Psi-nk})), there is a strong degeneracy between this parity violation modification and the spins, which prevented any constraints with the first LIGO observations. The coefficients $\delta\varpi$ and $\Delta^2$ only can be extracted as a linear combination, which is equivalent to that the coupling $\alpha$ and the spin parameters $\bm{S}_A$ are degenerate. However, taking the eccentricity into account, the modified phase includes many different combinations of $\delta\varpi$ and $\Delta^2$ at $\sim\mathcal{O}(e_0^2)$ and $\sim\mathcal{O}(e_0^4)$ orders, breaking this degeneracy between $\alpha$ and $\bm{S}_A$ in the eccentric binaries. Furthermore, the complete waveform includes the parts predicted by general relativity and DCS correction. The spin-orbit, spin-spin, and monopole-quadrupole coupling appear in 1.5PN, 2PN, and 2PN of the post-Newtonian waveform, respectively. However, these terms are not presented in this article for simplicity. This means that we can distinguish between the effects of spin and DCS correction. 

As a summary, we conclude that the final ready-to-use waveform is valid under four approximations. Among them, the first three, small-coupling, slowly-rotating, PN have been discussed in Section \ref{subsec:conserved-quantifies}. The last one is small initial eccentricity, which allows an analytic calculation of the frequency-domain evolution of eccentricity. We expand the expression up to the order of $\sim\mathcal{O}(e_0^4)$, which is valid for initial eccentricity $e_0\lesssim0.3$. Now, we make a rough estimate of the weakness of DCS correction. The small-coupling assumption requires $\zeta$ to be much smaller than $1$, and PN expansion requires the typical velocity $v$ to be much smaller than $1$. In conclusion, compared to the 2PN gravitational waveform, the correction is of order $\zeta\chi^2$, and compared to the Newtonian-order waveform, that is of order $\zeta v^4\chi^2$. It should be mentioned that the predictions for black hole spin magnitudes depend
upon the formation channel and assumptions about stellar
evolution and the low-spin black holes can be born by multiple mechanisms \cite{GWTC3,Fuller2019}. Therefore the waveform under slowly-rotating approximation should be still valid for such systems. However, for the significantly-spinning system, the waveform templates should be improved further.

\begin{figure}[h]
\centering
\includegraphics[width=0.5\linewidth]{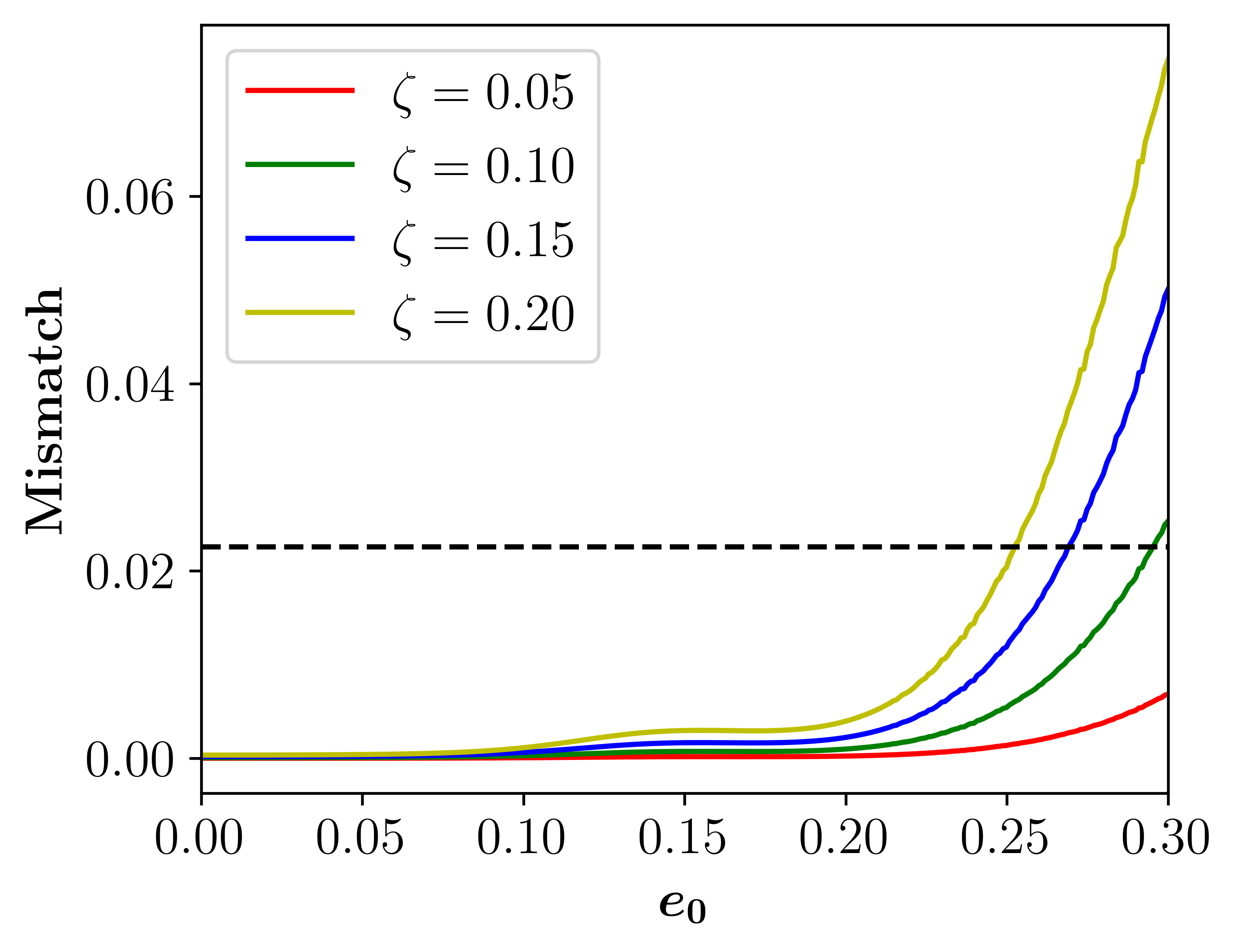}
\caption{The mismatch between eccentric waveforms between DCS theory and GR detected by advanced LIGO for different initial eccentricities and couplings. Here we consider a binary system with masses $m_1=98.4M_{\odot}$ and $m_2=57.2M_{\odot}$ (GW190521-like), corresponding ISCO frequency is $f_\text{ISCO}\approx28.17\text{Hz}$. In this calculation, the dimensionless spins of these two black holes are $0.15$ and $0.18$. The upper bound of coupling is taken as $0.2$, which is selected according to the small-coupling approximation, and satisfies the observational constraint given by \cite{Silva2021}. The inclination and azimuth angle of the detector and the polar and azimuth angle of the source are $30\deg$, $35\deg$, $40\deg$, and $50\deg$, respectively. The horizontal dashed line indicates mismatch $=0.0226$.}
\label{fig}
\end{figure}

Similar to Brans-Dicke and Einstein-dilaton-Gauss-Bonnet gravity, it is expected that including the eccentricity into the binary parameters enhances the distinguishment between DCS gravity and GR \cite{Moore2020}. To show that, we numerically calculate the mismatch between the frequency-domain waveforms in DCS theory and GR. The mismatch between waveforms $\tilde{h}_1(f)$ and $\tilde{h}_2(f)$ is defined by \cite{Khan2016,Knee2022}
\begin{equation}
\text{Mismatch}\equiv
1-\frac{\langle h_1,h_2\rangle}{\sqrt{\langle h_1,h_1\rangle\langle h_2,h_2\rangle}},
\end{equation}
with $\langle\cdot\rangle$ is the noise-weighted inner production of these two waveforms,
\begin{equation}
\langle h_1,h_2\rangle\equiv
4\ \text{Re}\int_{f_\text{low}}^{f_\text{high}}\frac{\tilde{h}_1(f)\tilde{h}_2^{*}(f)}{S_{n}(f)}df,
\end{equation}
where Re means the real part and $^*$ means the complex conjugate. $f_\text{high}$ and $f_\text{low}$ are the high and low-frequency cutoff of the integration. $f_\text{low}$ is usually set as $10$Hz for the second-generation ground-based detectors and $f_\text{high}$ should be the corresponding frequency for the innermost stable circular orbit (ISCO) of the binary, defined by $f_\text{ISCO}\equiv1/(6^{3/2}\pi m)$ \cite{ChangfuShi2022}. The PN approximation is no longer valid once the binary distance closes to the ISCO radius. In this work, $\tilde{h}_1$ is the eccentric waveforms in GR, with $\zeta=0$, and $\tilde{h}_2$ in DCS theory, with various coupling $\zeta\neq0$. $S_n(f)$ is the power spectral density of the GW detector. We consider a GW190521-like source detected by the advanced LIGO \cite{GWTC2}. The results of the mismatch are shown in Fig.\,\ref{fig}, which indicates that the mismatch is enhanced for higher-eccentricity binaries, in which the distinguishment between DCS theory and GR becomes greater. 

Usually, two waveforms are indistinguishable when their mismatch is smaller than $\chi_{k}^2(1-p)/2\rho^2$ \cite{Hannam2022}, where $\chi_k^2(1-p)$ is calculated from the cumulative distribution function of the chi-square distribution with $k$ degrees of freedom at probability $p$, and $\rho$ is the SNR. For spin-aligned eccentric binaries considered in this work, the degrees of freedom $k$ is $5$, and the mismatch criterion reduces to Mismatch $\leqslant4.62/\rho^2$. The SNR of GW190521 is about $14.3$ \footnote{https://gwosc.org/projects/} and thus the DCS modification is distinguishable when the mismatch reaches $0.0226$. The corresponding initial eccentricity from Fig.\,\ref{fig} is about $0.25$ for $\zeta=0.20$. This numerical result supports our statement that considering the eccentricity of the binaries brings more distinguishment between DCS gravity and GR. Therefore, our waveform model should be a reference to future GW observation and gravitational tests in the eccentric BBH systems.

\section{Conclusion and Discussion}
\label{sec:conslusion}
DCS theory \cite{ChernSimons2003,Alexander2009} is a parity-violating gravitational theory that has recently attracted more and more attention. Compared with other scalar-tensor gravity, this theory only modifies the non-spherically symmetric spacetime \cite{Grumiller2008,Konno2009,Yunes2009,Yagi2012}, such that only the gravitational radiation from binary rotating black holes encodes the distinctions between DCS and GR. In the PN framework, the DCS modifications to BBH motion, radiation, orbital secular evolution, and Fourier waveforms always enter 2PN-order correction \cite{Yagi2012pn,Yagi2012gw,ZhaoLi2023}. This is the main conclusion of our previous works \cite{ZhaoLi2023}, in which the quasi-circular orbits and waveforms have been fully investigated.

This article focuses on the non-precessing BBH systems with quasi-elliptic orbits. The motion is constrained on the orbital plane, thus the quasi-Keplerian parameterization firstly introduced for non-spinning binaries \cite{Damour1983} can be successfully extended to the DCS modification, which also induces the periastron-advance effects in Eq.\,(\ref{T-K}) and then the BBH orbits are no longer closed. Therefore, the BBH motion presents a doubly periodic structure, the azimuth angle of BBH passes through $(1+\beta)v$ while the true (or eccentric and mean) anomaly through $2\pi$. Two formal orbital elements, ``radial" semimajor $a_r$ and ``radial" eccentricity $e_r$, are introduced to describe the BBH motion [see Eq.\,(\ref{time-azimuth-integral-results})]. These two elements are not the true geometric quantifies but two independent parameters related to the conserved energy and OAM as shown in Eq.\,(\ref{elements-conserved-quantities}). 

Based on the PN description of the BBH motion, the scalar and gravitational waveforms are obtained through quadrupole formula \cite{ZhaoLi2023}, which are expressed in terms of true anomaly, rather than the azimuth angle. The waveform is shown as the linear combination of $\sin(nV)$ and $\cos(nV)$, with $n=1,2,3$ for Newtonian limit and $n=1,2,3,4,5$ for DCS modification [see Eqs.\,(\ref{vartheta-result}, \ref{Newtonian-polarization-in-V}, \ref{DCS-polarization-in-V})]. And again due to the periastron advance, the periodic behavior of waveform is modulated at a much lower frequency. From the results of the waveforms, the energy flux and OAM flux carried by scalar and tensor radiation are calculated as Eqs.\,(\ref{total-energy-flux}, \ref{total-angular-momentum-flux}). Combining the balance equation (\ref{balance-equation}), one can obtain the secular evolution of elements [see Eqs.\,(\ref{dx-dt}, \ref{der-dt})], which results in the orbits gradually becoming quasi-circular. The DCS modification changes the rate of circularization but does not influence the fact that the orbit will become circular. Although the time-domain solution of orbital frequency $x$ and eccentricity $e_r$ are unachievable, we can analytically solve the results of $x$ in terms of $e_r$, which indicate that the eccentricity decreases as frequency increases, which is shown in Eq.\,(\ref{dx-der}) and its solutions (\ref{x0-solution}, \ref{x1-solution}).

Due to the complicated form of secular evolution, the Fourier waveform cannot be fully obtained for arbitrary eccentricity. The small-eccentricity limit or post-circular approximation is in general adopted. Up to the fourth-order terms of initial eccentricity $e_0$ and the linear order of DCS coupling $\zeta$, the frequency-domain waveform is reported in Section \ref{sec:Fourier-waveform}. Specifically, the modified phase is shown in Eq.\,(\ref{Psi-nk}), and the amplitudes are shown in Appendix \ref{Appendix-D}, which completes the template construction. As shown in Ref.\,\cite{Nair2019,Perkins2021,ChangfuShi2022}, the observation of the BBH system without eccentricity can not give the physically meaningful constraint. However, similar to Brans-Dicke and Einstein-dilaton-Gauss-Bonnet gravity, it is expected that including the eccentricity into the binary parameters enhances the constraint on DCS effects \cite{Moore2020}. Especially, for the eccentric BBH system formed dynamically in densely populated environments \cite{Romero-Shaw2020,Gayathri2022,Romero-Shaw2022}, the DCS modification from GR is enhanced. As an example, the mismatch between eccentric waveforms in GR and DCS is calculated, and the result shown in Fig.\,\ref{fig}, means that a larger eccentricity will bring greater distinguishability. According to the above discussion, the BBH systems with significant eccentricity but low spin should be worth investigating further in future studies.

\begin{acknowledgments}
We would like to thank Aoxiang Jiang and Wei Liu for their helpful discussions and comments. This work is supported by the National Key R\&D Program of China Grant No.\,2022YFC2200100 and 2021YFC2203102, NSFC No.\,12273035 and 12325301 the Fundamental Research Funds for the Central Universities under Grant No. WK3440000004, and the science research grants from the China Manned Space Project with No.CMS-CSST-2021-B01. T. Z. is supported in part by the National Key Research and Development Program of China under Grant No.\,2020YFC2201503, the National Natural Science Foundation of China under Grant No.\,12275238 and No.\,11675143, the Zhejiang Provincial Natural Science Foundation of China under Grant No.\,LR21A050001 and LY20A050002,  and the Fundamental Research Funds for the Provincial Universities of Zhejiang in China under Grant No.\,RF-A2019015. T. L. is supported by NSFC No.\,12003008.
S. H. was supported by the National Natural Science Foundation of China under Grant No.~12205222, and by the Fundamental Research Funds for the Central Universities under Grant No.~2042022kf1062.
\end{acknowledgments}

\appendix
\section{\label{Appendix-A}Coefficients Involved in Eq.\,(\ref{chi1-solution})}
In \ref{subsec:eccentricity-Fourier}, we have derived the evolution of the orbital frequency with the changing eccentricity in the post-circular limits. Some coefficients used in Eq.\,(\ref{chi1-solution}) are listed as follows, 
\begin{equation}
\mathcal{P}^{(0)}_{0}
=-\mathcal{P}^{(0)}_{-24/19}
=\frac{325}{311296}\Delta^2
+\frac{41}{152}\delta\varpi,
\end{equation}
\begin{equation}
\begin{aligned}
\mathcal{P}^{(2)}_{0}&=
-\frac{3239925}{1798045696}\Delta^2
-\frac{408729}{877952}\delta\varpi,\qquad
\mathcal{P}^{(2)}_{-62/19}=
\frac{34620825}{12586319872}\Delta^2
-\frac{964753}{6145664}\delta\varpi,\Bigg.\\
\mathcal{P}^{(2)}_{-2}&=
\frac{40977425}{12586319872}\Delta^2
+\frac{10501763}{6145664}\delta\varpi,\qquad
\mathcal{P}^{(2)}_{-24/19}=
-\frac{7559825}{1798045696}\Delta^2
-\frac{953701}{877952}\delta\varpi,
\end{aligned}
\end{equation}
\begin{equation}
\begin{aligned}
\mathcal{P}^{(4)}_{0}&=\frac{4252453725}{5192755970048}\Delta^2+\frac{536463393}{2535525376}\delta\varpi,\qquad
\mathcal{P}^{(4)}_{-24/19}=-\frac{537569626725}{236270396637184}\Delta^2+\frac{141346710801}{115366404608}\delta\varpi,\Bigg.\\
\mathcal{P}^{(4)}_{-2}&=-\frac{408503949825}{72698583580672}\Delta^2-\frac{104692075347}{35497355264}\delta\varpi,\qquad
\mathcal{P}^{(4)}_{-62/19}=\frac{115045001475}{10385511940096}\Delta^2-\frac{3205874219}{5071050752}\delta\varpi,\Bigg.\\
\mathcal{P}^{(4)}_{-4}&=\frac{834586228575}{135011655221248}\Delta^2+\frac{314836109183}{65923659776}\delta\varpi,\qquad
\mathcal{P}^{(4)}_{-100/19}=-\frac{105761708325}{10385511940096}\Delta^2
-\frac{13342246281}{5071050752}\delta\varpi.
\end{aligned}
\end{equation}

\section{\label{Appendix-B}Coefficients Involved in Eq.\,(\ref{e-chi-coefficient})}
After obtaining the frequency represented in terms of evolving eccentricity, we inversely solve this relation to get the frequency-domain evolution of eccentricity, with some coefficients in Eq.\,(\ref{e-chi-coefficient}) shown as follows,
\begin{equation}
\begin{aligned}
\mathcal{S}^{(0)}_{0}
=-\mathcal{S}^{(0)}_{4/3}
=\frac{325}{294912}\Delta ^2
+\frac{41}{144}\delta\varpi
\end{aligned}
\end{equation}
\begin{equation}
\begin{aligned}
\mathcal{S}^{(2)}_{-19/9}&=
-\frac{1079975}{179306496}\Delta^2
-\frac{136243}{87552}\delta\varpi,\qquad
\mathcal{S}^{(2)}_{-7/9}=
\frac{5633725}{1255145472}\Delta^2
+\frac{49807}{204288}\delta\varpi,\Bigg.\\
\mathcal{S}^{(2)}_{0}&=
\frac{13338125}{3765436416}\Delta^2
+\frac{3366541}{1838592}\delta\varpi,\qquad
\mathcal{S}^{(2)}_{4/3}=
-\frac{1079975}{537919488}\Delta^2
-\frac{136243}{262656}\delta\varpi,
\end{aligned}
\end{equation}
\begin{equation}
\begin{aligned}
\mathcal{S}^{(4)}_{-38/9}&=\frac{81672082375}{1962330292224}\Delta^2+\frac{10303247315}{958169088}\delta\varpi,\qquad
\mathcal{S}^{(4)}_{-26/9}=-\frac{5196368177125}{178572056592384}\Delta^2-\frac{82850130257}{87193387008}\delta\varpi,\Bigg.\\
\mathcal{S}^{(4)}_{-19/9}&=-\frac{10507242925}{254376148992}\Delta^2-\frac{5841770863}{372621312}\delta\varpi,\qquad
\mathcal{S}^{(4)}_{-7/9}=\frac{18720868175}{763128446976}\Delta^2+\frac{165508661}{124207104}\delta\varpi,\Bigg.\\
\mathcal{S}^{(4)}_{0}&=\frac{1232639443225}{178572056592384}\Delta^2+\frac{455716402373}{87193387008}\delta\varpi,\qquad
\mathcal{S}^{(4)}_{4/3}=-\frac{5198125075}{1962330292224}\Delta^2-\frac{655763471}{958169088}\delta\varpi.
\end{aligned}
\end{equation}

\section{\label{Appendix-C}Coefficients Involved in Eq.\,(\ref{strain-exp})}
Up to the linear order in DCS couplings and the fourth order in the eccentricity, i.e., $\sim\mathcal{O}(\zeta)$ and $\sim\mathcal{O}(e_r^4)$, the time-domain waveform includes 26 modes denoted by different $n$'s and $k$'s. The non-vanishing modes included in Eq.\,(\ref{strain-exp}) with $k=2$ are
\begin{equation}
\begin{aligned}
A_{-1,2}&=\frac{1}{576}xe_r^3
\mathcal{F}_d^{(-)}
e^{2i\omega}(21+80\cdot\delta\varpi\cdot x^2),\\
A_{-2,2}&=\frac{1}{192}xe_r^4
\mathcal{F}_d^{(-)}
e^{2i\omega}(6+23\cdot\delta\varpi\cdot x^2),\\
A_{1,2}&=\frac{1}{64}xe_r\mathcal{F}_d^{(-)}
e^{2i\omega}[(24-13e_r^2)+(80+64e_r^2)\delta\varpi\cdot x ^2],\\
A_{2,2}&=-\frac{1}{96}\xi\mathcal{F}_d^{(-)}e^{2i\omega}
[(48-120e_r^2+69e_r^4)
+(32-376e_r^2-384e_r^4)\delta\varpi\cdot x^2],\\
A_{3,2}&=\frac{3}{64}xe_r
\mathcal{F}_d^{(-)}e^{2i\omega}
[-24+57e_r^2-(48-160e_r^2)\delta\varpi\cdot x^2],\\
A_{4,2}&=-\frac{1}{6}xe_r^2
\mathcal{F}_d^{(-)}e^{2i\omega}
[(12-30e_r^2)+(35-83e_r^2)\delta\varpi\cdot x^2],\\
A_{5,2}&=-\frac{25}{576}xe_r^3
\mathcal{F}_d^{(-)}e^{2i\omega}
(75+272\cdot\delta\varpi\cdot x^2),\\
A_{6,2}&=-\frac{3}{64}xe_r^4
\mathcal{F}_d^{(-)}e^{2i\omega}
(108+455\cdot\delta\varpi\cdot x^2),
\end{aligned}
\end{equation}
The other modes with $k=-2$ are
\begin{equation}
\begin{aligned}
A_{1,-2}&=\frac{1}{576}xe_r^3
\mathcal{F}_d^{(+)}e^{-2i\omega}
(21+80\cdot\delta\varpi\cdot x^2),\\
A_{2,-2}&=\frac{1}{192} xe_r^4
\mathcal{F}_d^{(+)}e^{-2i\omega}
(6+23\cdot\delta\varpi\cdot x^2),\\
A_{-1,-2}&=\frac{1}{64} xe_r
\mathcal{F}_d^{(+)}e^{-2i\omega}
[24-13e_r^2+(80+64e_r^2)\delta\varpi\cdot x^2],\\
A_{-2,-2}&=-\frac{1}{96} x
\mathcal{F}_d^{(+)}e^{-2i\omega}
[(48-120e_r^2+69e_r^4)
+(32-384e_r^2-276e_r^4)\delta\varpi\cdot x^2],\\
A_{-3,-2}&=-\frac{3}{64} xe_r
\mathcal{F}_d^{(+)}e^{-2i\omega}
[(24-57e_r^2)+(48-160e_r^2)\delta\varpi\cdot x^2],\\
A_{-4,-2}&=-\frac{1}{6} xe_r^2
\mathcal{F}_d^{(+)}e^{-2i\omega}
[(12-30e_r^2)+(35-83e_r^2)\delta\varpi\cdot x^2],\\
A_{-5,-2}&=-\frac{25}{576} xe_r^3
\mathcal{F}_d^{(+)}e^{-2i\omega}
(75+272\cdot\delta\varpi\cdot x^2),\\
A_{-6,-2}&=-\frac{3}{64} xe_r^4
\mathcal{F}_d^{(+)}e^{-2i\omega}
(108+455\cdot\delta\varpi\cdot x^2).
\end{aligned}
\end{equation}
The others vanish in the considered order, i.e.,
\begin{equation}
A_{0,2}=A_{-3,2}=A_{-4,2}=A_{-5,2}
=A_{-6,2}=A_{0,-2}=A_{3,-2}=A_{4,-2}
=A_{5,-2}=A_{6,-2}=0.
\end{equation}
We recall that $x$ is the dimensionless orbital frequency, defined by $x\equiv(m\Omega)^{2/3}$ and $\omega$ is the azimuth coordinate of the observer. Additionally, we define a new symbol $\mathcal{F}_d^{(\pm)}$ in terms of the pattern functions $F_{+,\times}$ and the inclination of the observers in the following way, 
\begin{equation}
\mathcal{F}_d^{(\pm)}
=(1+\cos^2\iota)F_{+}\pm2i\cos\iota F_{\times}.
\end{equation}

\section{\label{Appendix-D}Detailed Calculation of Frequency-Domain Waveform}
This appendix presents the detailed calculation of the frequency-domain waveform from Eq.\,(\ref{SPA}) through the method shown in Sec \ref{overview-Fourier}. The stationary points in integration (\ref{SPA}) are determined by $f=(n+k\beta_{nk})F_{nk}$, where the precession rate is
\begin{equation}
\label{beta-F}
\beta(F)=\delta\varpi\cdot\frac{\tilde{u}^{4}}{\nu^{4/5}}
\left\{1+2e_0^2\chi^{-19/9}+e_0^4
\left[\frac{3323}{456}\chi^{-19/9}
-\frac{1955}{456}\chi^{-38/9}\right]\right\}.
\end{equation}
Here we defined a new dimensionless frequency
$\tilde{u}\equiv(2\pi\mathcal{M}F)^{1/3}$, and the chirp mass $\mathcal{M}\equiv m\nu^{3/5}$. Therefore, the solution of stationary frequency, corresponding to Eq.\,(\ref{stationary-point-equation-overview}), is
\begin{equation}
\label{stationary-point}
F_{nk}=F_{nk}(f)
\approx\frac{f}{n}
\left\{1+\frac{k}{n}\delta\varpi
\frac{\tilde{u}_f^{4}}{\nu^{4/5}}
\left[1-2e_0^2\chi_{f}^{-19/9}
+e_0^4\left(\frac{1955}{456}\chi_{f}^{-38/9}
-\frac{3323}{456}\chi_{f}^{-19/9}\right)\right]\right\},
\end{equation}
where $\tilde{u}_f\equiv(2\pi\mathcal{M}f/n)^{1/3}$ and $\chi_{f}\equiv(1/n)(f/F_0)$. Using formula (\ref{derivative-F}), the first and second order derivatives of orbital frequency $F$ are given by
\begin{equation}
\label{dot-F-nk}
\begin{aligned}
\frac{5}{48}\pi\mathcal{M}^2\dot{F}&=\tilde{u}^{11}
\left\{1+\frac{157}{24}e_0^2\chi^{-19/9}
+e_0^4\chi^{-19/9}
\left[-\frac{107891}{21888}
+\frac{521711}{21888}\chi^{-19/9}\right]\right\}\\
&+\frac{\tilde{u}^{15}}{\nu^{4/5}}\Bigg\{
\left[\frac{25}{24576}\Delta^2
+\frac{10}{3}\delta\varpi\right]\\
&+e_0^2\chi^{-19/9}
\left[\left(\frac{2975}{3538944}
+\frac{51025}{3538944}\chi^{-4/3}\right)\Delta^2
+\left(\frac{50011}{1728}
+\frac{6437}{1728}\chi^{-4/3}\right)\delta\varpi\right]\\
&+e_0^4\chi^{-19/9}\Bigg[
\left(\frac{9885925}{3227516928}+\frac{1640489075}{22592618496}\chi^{-4/3}+\frac{147769375}{5648154624}\chi^{-19/9}-\frac{35064575}{1613758464}\chi^{-31/9}\right)\Delta^2\\
&\qquad\qquad+\left(\frac{166186553}{1575936}+\frac{339138997}{11031552}\chi^{-4/3}+\frac{83973067}{5515776}\chi^{-19/9}-\frac{4423531}{787968}\chi^{-31/9}\right)\delta\varpi\Bigg]\Bigg\},
\end{aligned}
\end{equation}
and
\begin{equation}
\label{ddot-F-nk}
\begin{aligned}
\frac{25}{16896}\pi\mathcal{M}^3\ddot{F}&=\tilde{u}^{11}
\left\{1+\frac{7379}{792}e_0^2\chi^{-19/9}
+e_0^4\chi^{-19/9}\left[\frac{24520417}{722304}+\frac{315385}{22572}\chi^{-19/9}\right]\right\}\\
&+\frac{\tilde{u}^{15}}{\nu^{4/5}}\Bigg\{
\left[\frac{325}{135168}\Delta^2
+\frac{260}{33}\delta\varpi\right]\\
&+e_0^2\chi^{-19/9}
\left[\left(\frac{1564975}{116785152}+\frac{2398175}{116785152}\chi^{-4/3}\right)\Delta^2
+\left(\frac{5173769}{57024}+\frac{302539}{57024}\chi^{-4/3}\right)\delta\varpi\right]\\
&+e_0^4\chi^{-19/9}\Bigg[\left(\frac{5200411925}{106508058624}+\frac{77102986525}{745556410368}\chi^{-4/3}+\frac{752542375}{23298637824}\chi^{-19/9}+\frac{102500125}{1664188416}\chi^{-31/9}\right)\Delta^2\\
&\qquad\qquad+\left(\frac{17192434387}{52005888}+\frac{15939532859}{364041216}\chi^{-4/3}+\frac{2594060795}{11376288}\chi^{-19/9}+\frac{12930785}{812592}\chi^{-31/9}\right)\delta\varpi\Bigg]\Bigg\},
\end{aligned}
\end{equation}
respectively. Thus, using Eq.\,(\ref{mean-anomaly-overview}), the mean anomaly is integrated as
\begin{equation}
\label{ell-F}
\begin{aligned}
\ell(F)&=\ell_c-\frac{1}{32}\tilde{u}^{-5}
\left\{1-\frac{785}{272}e_0^2\chi^{-19/9}
-e_0^4\chi^{-19/9}\left[\frac{2608555}{248064}-\frac{5222765}{386688}\chi^{-19/9}\right]\right\}\\
&-\frac{1}{32}\frac{\tilde{u}^{-1}}{\nu^{4/5}}\Bigg\{\left[-\frac{125}{24576}\Delta^2-\frac{50}{3}\delta\varpi\right]\\
&+e_0^2\chi^{-19/9}
\left[\left(\frac{220625}{25952256}-\frac{255125}{40108032}\chi^{-4/3}\right)\Delta^2
+\left(\frac{126745}{12672}-\frac{32185}{19584}\chi^{-4/3}\right)\delta\varpi\right]\\
&+e_0^4\chi^{-19/9}\Bigg[
\Bigg(\frac{733136875}{23668457472}-\frac{8202445375}{256049676288}\chi^{-4/3}-\frac{1099890625}{19297861632}\chi^{-19/9}+\frac{1697398625}{28509732864}\chi^{-31/9}\Bigg)\Delta^2\\
&\qquad\qquad+\left(\frac{421173635}{11556864}-\frac{1695694985}{125024256}\chi^{-4/3}-\frac{2685294025}{75382272}\chi^{-19/9}+\frac{214133365}{13920768}\chi^{-31/9}\right)\delta\varpi\Bigg]\Bigg\}.
\end{aligned}
\end{equation}
The first and second order derivatives are obtained from Eq.\,(\ref{derivative-mean-anomaly-overview}), represented as
\begin{equation}
\label{dot-ell-F}
\dot{\ell}(F)=2\pi F,
\end{equation}
and
\begin{equation}
\label{ddot-ell-F}
\begin{aligned}
\ddot{\ell}(F)
&=\frac{96}{5}\frac{\tilde{u}^{11}}{\mathcal{M}^2}
\left\{1+\frac{157}{24}e_0^2\chi^{-19/9}
+e_0^4\left[-\frac{107891}{21888}\chi^{-38/9}
+\frac{521711}{21888}\chi^{-19/9}\right]\right\}\\
&+\frac{96}{5}\frac{1}{\mathcal{M}^2}
\frac{\tilde{u}^{15}}{\nu^{4/5}}\Bigg\{
\left[\frac{25}{24576}\Delta^2
+\frac{10}{3}\delta\varpi\right]\\
&+e_0^2\chi^{-19/9}
\Bigg[\left(\frac{2975}{3538944}+\frac{51025}{3538944}\chi^{-4/3}\right)\Delta^2+\left(\frac{50011}{1728}+\frac{6437}{1728}\chi^{-4/3}\right)\delta\varpi\Bigg]\\
&+e_0^4\chi^{-19/9}\Bigg[\left(\frac{9885925}{3227516928}+\frac{1640489075}{22592618496}\chi^{-4/3}+\frac{147769375}{5648154624}\chi^{-19/9}-\frac{35064575}{1613758464}\chi^{-31/9}\right)\Delta^2\\
&\qquad\qquad+\left(\frac{166186553}{1575936}+\frac{339138997}{11031552}\chi^{-4/3}+\frac{83973067}{5515776}\chi^{-19/9}-\frac{4423531}{787968}\chi^{-31/9}\right)\delta\varpi\Bigg]\Bigg\}.
\end{aligned}
\end{equation}
Substituting Eq.\,(\ref{stationary-point}) into Eqs.\,(\ref{ell-F}, \ref{dot-ell-F}, \ref{ddot-ell-F}), we obtain the value of mean anomaly and its derivatives at the stationary points, which are shown as
\begin{equation}
\label{ell-nk}
\begin{aligned}
\ell_{nk}&=\ell_c-\frac{1}{32}\tilde{u}_f^{-5}
\left\{1-\frac{785}{272}e_0^2\chi_f^{-19/9}
+e_0^4\left[\frac{5222765}{386688}\chi_f^{-38/9}
-\frac{2608555}{248064}\chi_f^{-19/9}\right]\right\}\\
&-\frac{1}{32}\frac{\tilde{u}_f^{-1}}{\nu^{4/5}}\Bigg\{\left[-\frac{125}{24576}\Delta^2+\left(\frac{5}{3}\frac{k}{n}-\frac{50}{3}\right)\delta\varpi\right]\\
&+e_0^2\chi_f^{-19/9}
\left[\left(\frac{220625}{25952256}-\frac{255125}{40108032}\chi_f^{-4/3}\right)\Delta^2+\left(-\frac{545}{72}\frac{k}{n}+\frac{126745}{12672}-\frac{32185}{19584}\chi_f^{-4/3}\right)\delta\varpi\right]\\
&+e_0^4\chi_f^{-19/9}\Bigg[
\left(\frac{733136875}{23668457472}-\frac{8202445375}{256049676288}\chi_f^{-4/3}-\frac{1099890625}{19297861632}\chi_f^{-19/9}+\frac{1697398625}{28509732864}\chi_f^{-31/9}\right)\Delta^2\\
&\qquad+\Bigg(\left(\frac{421173635}{11556864}
-\frac{1811035}{65664}\frac{k}{n}\right)-\frac{1695694985}{125024256}\chi_f^{-4/3}\\
&\qquad-\left(\frac{2685294025}{75382272}-\frac{3321725}{65664}\frac{k}{n}\right)\chi_f^{-19/9}+\frac{214133365}{13920768}\chi_f^{-31/9}\Bigg)\delta\varpi\Bigg]\Bigg\},
\end{aligned}
\end{equation}
\begin{equation}
\label{dot-ell-nk}
\dot{\ell}_{nk}=\dot{\ell}(F_{nk})=
\frac{2\pi f}{n}\left\{1+\frac{k}{n}\delta\varpi
\frac{\tilde{u}_f^4}{\nu^{4/5}}
\left[1+2e_0^2\chi_f^{-19/9}
+e_0^4\chi_f^{-19/9}\left(\frac{3323}{456}
-\frac{1955}{456}\chi_f^{-19/9}\right)\right]\right\},
\end{equation}
and
\begin{equation}
\label{ddot-ell-nk}
\begin{aligned}
\ddot{\ell}_{nk}
&=\frac{96}{5}\frac{\tilde{u}_f^{11}}{\mathcal{M}^2}
\left\{1+\frac{157}{24}e_0^2\chi_f^{-19/9}
+e_0^4\chi_f^{-19/9}\left[\frac{521711}{21888}-\frac{107891}{21888}\chi_f^{-19/9}\right]\right\}\\
&+\frac{96}{5}\frac{1}{\mathcal{M}^2}
\frac{\tilde{u}_f^{15}}{\nu^{4/5}}\Bigg\{
\left[\frac{25}{24576}\Delta^2
+\left(\frac{10}{3}-\frac{11}{3}\frac{k}{n}\right)\delta\varpi
\right]\\
&+e_0^2\cdot\chi_f^{-19/9}
\Bigg[\left(\frac{2975}{3538944}+\frac{51025}{3538944}\chi_f^{-4/3}\right)\Delta^2+\left(\frac{50011}{1728}-\frac{1891}{108}\frac{k}{n}+\frac{6437}{1728}\chi_f^{-4/3}\right)\delta\varpi\Bigg]\\
&+e_0^4\chi_f^{-19/9}\Bigg[\left(\frac{9885925}{3227516928}+\frac{1640489075}{22592618496}\chi_f^{-4/3}+\frac{147769375}{5648154624}\chi_f^{-19/9}-\frac{35064575}{1613758464}\chi_f^{-31/9}\right)\Delta^2\\
&\qquad+\Bigg(\left(\frac{166186553}{1575936}-\frac{6283793}{98496}\frac{k}{n}\right)+\frac{339138997}{11031552}\chi_f^{-4/3}\\
&\qquad+\left(\frac{83973067}{5515776}-\frac{1451887}{196992}\frac{k}{n}\right)\chi_f^{-19/9}-\frac{4423531}{787968}\chi_f^{-31/9}\Bigg)\delta\varpi\Bigg]\Bigg\}.
\end{aligned}
\end{equation}
Similarly, the precession rate $\beta$ has been written as the function of orbital frequency, i.e., $\beta=\beta(F)$ (\ref{beta-F}). Its derivatives are given by Eq.\,(\ref{derivative-beta}), 
\begin{equation}
\label{dot-beta-F}
\dot{\beta}(F)=\delta\varpi
\cdot\frac{128}{5}\frac{1}{\mathcal{M}}
\frac{\tilde{u}^{12}}{\nu^{4/5}}
\left[1+\frac{43}{8}e_0^2\chi^{-19/9}
+e_0^4\chi^{-19/9}\left(\frac{142889}{7296}
-\frac{23873}{7296}\chi^{-19/9}\right)\right],
\end{equation}
and
\begin{equation}
\label{ddot-beta-F}
\ddot{\beta}(F)
=\delta\varpi\cdot\frac{49152}{25}\frac{1}{\mathcal{M}^2}
\frac{\tilde{u}^{20}}{\nu^{4/5}}
\left[1+\frac{2615}{288}e_0^2\chi^{-19/9}
+e_0^4\chi^{-19/9}\left(\frac{8689645}{262656}
+\frac{389275}{32832}\chi^{-19/9}\right)\right].
\end{equation}
The corresponding values at the stationary points are
\begin{equation}
\label{beta-nk}
\beta_{nk}=\delta\varpi\cdot\frac{\tilde{u}_f^{4}}{\nu^{4/5}}
\left\{1+2e_0^2\chi_f^{-19/9}+e_0^4
\chi_f^{-19/9}\left[\frac{3323}{456}
-\frac{1955}{456}\chi_f^{-19/9}\right]\right\},
\end{equation}
\begin{equation}
\label{dot-beta-nk}
\dot{\beta}_{nk}
=\delta\varpi\cdot\frac{128}{5}\frac{1}{\mathcal{M}}
\frac{\tilde{u}_f^{12}}{\nu^{4/5}}
\left\{1+\frac{43}{8}e_0^2\chi_f^{-19/9}
+e_0^4\chi_f^{-19/9}\left[\frac{142889}{7296}
-\frac{23873}{7296}\chi_f^{-19/9}\right]\right\},
\end{equation}
and
\begin{equation}
\label{ddot-beta-nk}
\ddot{\beta}_{nk}
=\delta\varpi\cdot\frac{49152}{25}\frac{1}{\mathcal{M}^2}
\frac{\tilde{u}_f^{20}}{\nu^{4/5}}
\left[1+\frac{2615}{288}e_0^2\chi_f^{-19/9}
+e_0^4\chi_f^{-19/9}\left(\frac{8689645}{262656}
+\frac{389275}{32832}\chi_f^{-19/9}\right)\right],
\end{equation}
respectively. Finally, from Eq.\,(\ref{time}), the time function is 
\begin{equation}
\label{t-F}
\begin{aligned}
t(F)&=t_c-\frac{5}{256}\mathcal{M}\tilde{u}^{-8}
\left\{1-\frac{157}{43}e_0^2\chi^{-19/9}
+e_0^4\left[\frac{1044553}{56544}\chi^{-38/9}
-\frac{521711}{39216}\chi^{-19/9}\right]\right\}\\
&-\frac{5}{256}\mathcal{M}
\frac{\tilde{u}^{-4}}{\nu^{4/5}}\Bigg\{\left[-\frac{25}{12288}\Delta^2-\frac{20}{3}\delta\varpi\right]\\
&+e_0^2\chi^{-19/9}\left[\left(\frac{44125}{4571136}-\frac{51025}{6340608}\chi^{-4/3}\right)\Delta^2+\left(\frac{25349}{2232}-\frac{6437}{3096}\chi^{-4/3}\right)\delta\varpi\right]\\
&+e_0^4\chi^{-19/9}\Bigg[
\left(\frac{146627375}{4168876032}-\frac{1640489075}{40478441472}\chi^{-4/3}-\frac{8799125}{117669888}\chi^{-19/9}+\frac{339479725}{4168876032}\chi^{-31/9}\right)\Delta^2\\
&\qquad\qquad+\Bigg(\frac{84234727}{2035584}-\frac{339138997}{19764864}\chi^{-4/3}-\frac{107411761}{2298240}\chi^{-19/9}+\frac{42826673}{2035584}\chi^{-31/9}\Bigg)\delta\varpi\Bigg]\Bigg\},
\end{aligned}
\end{equation}
and the stationary-point value is thus
\begin{equation}
\label{t-nk}
\begin{aligned}
t_{nk}&=t_c-\frac{5}{256}\mathcal{M}\tilde{u}_f^{-8}
\left\{1-\frac{157}{43}e_0^2\chi_f^{-19/9}
+e_0^4\left[\frac{1044553}{56544}\chi_f^{-38/9}
-\frac{521711}{39216}\chi_f^{-19/9}\right]\right\}\\
&-\frac{5}{256}\mathcal{M}
\frac{\tilde{u}_f^{-4}}{\nu^{4/5}}\Bigg\{
\left[-\frac{25}{12288}\Delta^2
+\left(\frac{8}{3}\frac{k}{n}
-\frac{20}{3}\right)\delta\varpi\right]\\
&+e_0^2\chi_f^{-19/9}
\Bigg[\left(\frac{44125}{4571136}-\frac{51025}{6340608}\chi_f^{-4/3}\right)\Delta^2+\left(\frac{25349}{2232}-\frac{109}{9}\frac{k}{n}-\frac{6437}{3096}\chi_f^{-4/3}\right)\delta\varpi\Bigg]\\
&+e_0^4\chi_f^{-19/9}\Bigg[
\Bigg(\frac{146627375}{4168876032}-\frac{1640489075}{40478441472}\chi_f^{-4/3}-\frac{8799125}{117669888}\chi_f^{-19/9}+\frac{339479725}{4168876032 }\chi_f^{-31/9}\Bigg)\Delta^2\\
&\qquad\qquad+\Bigg(\left(\frac{84234727}{2035584}-\frac{362207}{8208}\frac{k}{n}\right)-\frac{339138997}{19764864}\chi_f^{-4/3}\\
&\qquad\qquad-\left(\frac{107411761}{2298240}-\frac{664345}{8208}\frac{k}{n}\right)\chi_f^{-19/9}+\frac{42826673}{2035584}\chi_f^{-31/9}\Bigg)
\delta\varpi\Bigg]\Bigg\}.
\end{aligned}
\end{equation}
Using Eqs.\,(\ref{t-nk}, \ref{beta-nk}, \ref{ell-nk}), the final modified phase defined in Eqs.\,(\ref{strain-frequency-overview}) or (\ref{modified-phase-overview}) is given in the main text.

\section{\label{Appendix-E}The Explicit Results of the Modified Amplitudes}
At the end of the main text, we have presented the amplitude and phase of the Fourier waveform. The modified amplitudes are listed in this appendix. We first define an overall coefficient as  
\begin{equation}
\label{D1}
\mathcal{A}_{n}^{(0)}\equiv-\sqrt{\frac{5}{96}}
\pi^{-2/3}\frac{\mathcal{M}^{5/6}}{R}f^{-7/6}
\left(\frac{n}{2}\right)^{2/3}.
\end{equation}
The non-vanishing modes in GR are
\begin{equation}
\begin{aligned}
\bar{\mathcal{A}}_{1,2}
&=-\frac{7}{96}\mathcal{A}_1^{(0)}
\mathcal{F}_{d}^{(-)}e^{2i\omega}e_0^3\chi_f^{-19/6},
\qquad
\bar{\mathcal{A}}_{2,2}
=-\frac{1}{16}\mathcal{A}_2^{(0)}
\mathcal{F}_{d}^{(-)}e^{2i\omega}e_0^4\chi_f^{-38/9},\\
\bar{\mathcal{A}}_{1,-2}
&=\mathcal{A}_1^{(0)}\mathcal{F}_{d}^{(+)}e^{-2i\omega}
\left\{-\frac{3}{4}e_0\chi_f^{-19/18}
+e_0^3\left[\frac{10277}{2432}\chi_f^{-19/6}
-\frac{3323}{2432}\chi_f^{-19/18}\right]\right\},\\
\bar{\mathcal{A}}_{2,-2}
&=\mathcal{A}_2^{(0)}\mathcal{F}_{d}^{(+)}e^{-2i\omega}
\left\{1-\frac{277}{48}e_0^2\chi_f^{-19/9}
+e_0^4\left[\frac{3260071}{87552}\chi_f^{-38/9}
-\frac{920471}{43776}\chi_f^{-19/9}\right]\right\},\\
\bar{\mathcal{A}}_{3,-2}
&=\mathcal{A}_3^{(0)}\mathcal{F}_{d}^{(+)}e^{-2i\omega}
\left\{\frac{9}{4}e_0\chi_f^{-19/18}
+e_0^3\left[\frac{9969}{2432}\chi_f^{-19/18}
-\frac{40863}{2432}\chi_f^{-19/6}\right]\right\},\\
\bar{\mathcal{A}}_{4,-2}
&=\mathcal{A}_4^{(0)}\mathcal{F}_{d}^{(+)}e^{-2i\omega}
\left\{4e_0^2\chi_f^{-19/9}
+e_0^4\left[\frac{3323}{228}\chi_f^{-19/9}
-\frac{1431}{38}\chi_f^{-38/9}\right]\right\},\\
\bar{\mathcal{A}}_{5,-2}
&=\frac{625}{96}\mathcal{A}_5^{(0)}
\mathcal{F}_{d}^{(+)}e^{-2i\omega}
e_0^3\chi_f^{-19/6},\qquad
\bar{\mathcal{A}}_{6,-2}
=\frac{81}{8}\mathcal{A}_6^{(0)}
\mathcal{F}_{d}^{(+)}e^{-2i\omega}
e_0^4\chi_f^{-38/9}.
\end{aligned}
\end{equation}
The non-vanishing modes in DCS modification are
\begin{equation}
\begin{aligned}
\delta{\mathcal{A}}_{1,2}
&=\mathcal{A}_1^{(0)}
\mathcal{F}_{d}^{(-)}e^{2i\omega}e_0^3\chi_f^{-19/6}\frac{\tilde{u}_f^{4}}{\nu^{4/5}}\Bigg\{\left(\frac{875}{3145728}-\frac{2275}{9437184}\chi_f^{-4/3}\right)\Delta^2\\
&\qquad\qquad-\left(\frac{287}{4608}+\frac{15941}{23040}\chi_f^{-4/3}+\frac{112}{15}\chi_f^{3/2}(2\pi\mathcal{M}F_0)^{5/3}\ell_c\right)\delta\varpi\Bigg\},\\
\delta{\mathcal{A}}_{2,2}
&=\mathcal{A}_2^{(0)}
\mathcal{F}_{d}^{(-)}e^{2i\omega}e_0^4\chi_f^{-38/9}\frac{\tilde{u}_f^{4}}{\nu^{4/5}}\Bigg\{\left(\frac{725}{2359296}-\frac{325}{1179648}\chi_f^{-4/3}\right)\Delta^2\\
&\qquad\qquad+\left(-\frac{371}{960}-\frac{41}{576}\chi_f^{-4/3}+\frac{16}{5}\chi_{f}^{-23/9}(2\pi\mathcal{M}F_0)^{5/3}\ell_c\right)\delta\varpi\Bigg\},
\end{aligned}
\end{equation}
\begin{equation}
\begin{aligned}
\delta{\mathcal{A}}_{1,-2}
&=\mathcal{A}_1^{(0)}\mathcal{F}_{d}^{(+)}e^{-2i\omega}\chi_f^{-19/18}\frac{\tilde{u}_f^{4}}{\nu^{4/5}}\Bigg\{\Bigg[e_0\left(\frac{475}{393216}-\frac{325}{393216}\chi_f^{-4/3}\right)\\
&+e_0^3\left(\frac{83075}{37748736}-\frac{13338125}{5020581888}\chi_f^{-4/3}-\frac{1034275}{88080384}\chi_f^{-19/9}+\frac{3340025}{239075328}\chi_f^{-31/9}\right)\Bigg]\Delta^2\\
&+\Bigg[e_0\left(\frac{623}{320}-\frac{41}{192}\chi_f^{-4/3}-\frac{384}{5}\chi_f^{11/18}(2\pi\mathcal{M}F_0)^{5/3}\ell_c\right)\\
&+e_0^3\Bigg(\frac{2070229}{583680}-\frac{3366541}{2451456}\chi_f^{-4/3}-\frac{1587076063}{69457920}\chi_f^{-19/9}+\frac{421357}{116736}\chi_f^{-31/9}\\
&+\left(\frac{67768}{285}\chi_f^{-3/2}-\frac{13292}{95}\chi_f^{11/18}\right)(2\pi\mathcal{M}F_0)^{5/3}\ell_c\Bigg)\Bigg]\delta\varpi\Bigg\},
\end{aligned}
\end{equation}
\begin{equation}
\begin{aligned}
\delta{\mathcal{A}}_{2,-2}
&=\mathcal{A}_2^{(0)}\mathcal{F}_{d}^{(+)}e^{-2i\omega}\frac{\tilde{u}_f^{4}}{\nu^{4/5}}\Bigg\{\Bigg[-\frac{25}{49152}+e_0^2\left(\frac{40175}{3538944}\chi_f^{-19/9}-\frac{90025}{7077888}\chi_f^{-31/9}\right)\\
&+e_0^4\left(\frac{133501525}{3227516928}\chi_f^{-19/9}-\frac{2894366075}{45185236992}\chi_f^{-31/9}-\frac{11491459625}{90370473984}\chi_f^{-38/9}+\frac{1059523075}{6455033856}\chi_f^{-50/9}\right)\Bigg]\Delta^2\\
&+\Bigg[
-\frac{29}{15}+\frac{256}{5}\chi_f^{5/3}(2\pi\mathcal{M}F_0)^{5/3}\ell_c
+e_0^2\left(\frac{3680737}{293760}\chi_f^{-19/9}-\frac{11357}{3456}\chi_f^{-31/9}-\frac{7448}{45}\chi_f^{-4/9}(2\pi\mathcal{M}F_0)^{5/3}\ell_c\right)\\
&+e_0^4\Bigg(\frac{643741529}{14100480}\chi_f^{-19/9}-\frac{598353517}{22063104}\chi_f^{-31/9}-\frac{686542948523}{5231278080}\chi_f^{-38/9}+\frac{133662911}{3151872}\chi_f^{-50/9}\\
&+\left(\frac{157387}{135}\chi_f^{-23/9}-\frac{162827}{270}\chi_f^{-4/9}\right)(2\pi\mathcal{M}F_0)^{5/3}\ell_c\Bigg)\Bigg]\delta\varpi\Bigg\},
\end{aligned}
\end{equation}
\begin{equation}
\begin{aligned}
\delta{\mathcal{A}}_{3,-2}
&=\mathcal{A}_3^{(0)}
\mathcal{F}_{d}^{(+)}e^{-2i\omega}\chi_f^{-19/18}\frac{\tilde{u}_f^{4}}{\nu^{4/5}}\Bigg\{\Bigg[e_0\left(-\frac{475}{131072}+\frac{325}{131072}\chi_f^{-4/3}\right)\\
&+e_0^3\left(-\frac{2759}{960}
+\frac{41}{64}\chi_f^{-4/3}+\frac{1496275}{29360128}\chi_f^{-19/9}-\frac{4426825}{79691776}\chi_f^{-31/9}\right)\Bigg]\Delta^2\\
&+\Bigg[e_0\left(-\frac{9168157}{1751040}-\frac{41}{192}\chi_f^{-4/3}+\frac{384}{5}\chi_f^{11/18}(2\pi\mathcal{M}F_0)^{5/3}\ell_c\right)\\
&+e_0^3\Bigg(\frac{2070229}{583680}+\frac{3366541}{817152}\chi_f^{-4/3}+\frac{1981498061}{69457920}\chi_f^{-19/9}-\frac{558461}{38912}\chi_f^{-31/9}\\
&+\left(\frac{13292}{95}\chi_f^{11/18}-\frac{107896}{285}\chi_f^{-3/2}\right)(2\pi\mathcal{M}F_0)^{5/3}\ell_c\Bigg)\Bigg]\delta\varpi\Bigg\},
\end{aligned}
\end{equation}
\begin{equation}
\begin{aligned}
\delta{\mathcal{A}}_{4,-2}
&=\mathcal{A}_4^{(0)}
\mathcal{F}_{d}^{(+)}e^{-2i\omega}\chi_f^{-19/9}\frac{\tilde{u}_f^{4}}{\nu^{4/5}}\Bigg\{\Bigg[e_0^2\left(-\frac{25}{2304}+\frac{325}{36864}\chi_f^{-4/3}\right)\\
&+e_0^4\left(-\frac{83075}{2101248}+\frac{10448975}{235339776}\chi_f^{-4/3}+\frac{678425}{4358144}\chi_f^{-19/9}-\frac{51675}{311296}\chi_f^{-31/9}\right)\Bigg]\Delta^2\\
&+\Bigg[e_0^2\left(-\frac{101}{30}+\frac{41}{18}\chi_f^{-4/3}+\frac{512}{5}\chi_f^{-4/9}(2\pi\mathcal{M}F_0)^{5/3}\ell_c\right)\\
&+e_0^4\Bigg(-\frac{335623}{27360}+\frac{2160121}{114912}\chi_f^{-4/3}+\frac{39932437}{813960}\chi_f^{-19/9}-\frac{6519}{152}\chi_f^{-31/9}\\
&+\left(\frac{106336}{285}\chi_f^{-4/9}-\frac{602032}{855}\chi_f^{-23/9}\right)(2\pi\mathcal{M}F_0)^{5/3}\ell_c\Bigg)\Bigg]\delta\varpi\Bigg\},
\end{aligned}
\end{equation}
\begin{equation}
\label{D8}
\begin{aligned}
\delta{\mathcal{A}}_{5,-2}
&=\mathcal{A}_5^{(0)}
\mathcal{F}_{d}^{(+)}e^{-2i\omega}e_0^3\chi_f^{-19/6}\frac{\tilde{u}_f^{4}}{\nu^{4/5}}\Bigg\{
\left(-\frac{78125}{3145728}+\frac{203125}{9437184}\chi_f^{-4/3}\right)\Delta^2\\
&+\left(-\frac{16025}{4608}+\frac{25625}{4608}\chi_f^{-4/3}+\frac{400}{3}\chi_f^{-3/2}(2\pi\mathcal{M}F_0)^{5/3}\ell_c\right)\delta\varpi\Bigg\},\\
\delta{\mathcal{A}}_{6,-2}
&=\mathcal{A}_6^{(0)}
\mathcal{F}_{d}^{(+)}e^{-2i\omega}e_0^4\chi_f^{-38/9}\frac{\tilde{u}_f^{4}}{\nu^{4/5}}\Bigg\{\left(-\frac{6525}{131072}+\frac{2925}{65536}\chi_f^{-4/3}\right)\Delta^2\\
&+\left(-\frac{63}{20}+\frac{369}{32}\chi_f^{-4/3}+\frac{864}{5}\chi_f^{-23/9}(2\pi\mathcal{M}F_0)^{5/3}\ell_c\right)\delta\varpi\Bigg\}.
\end{aligned}
\end{equation}

\bibliographystyle{apsrev4-2}
\bibliography{reference}
\end{document}